\documentclass[a4paper,14pt]{article}
\pdfoutput=1 

\usepackage{jheppub} 

\usepackage[T1]{fontenc} 
\usepackage{csquotes}
\usepackage{todonotes}
\usepackage{slashed}
\usepackage{cancel} 
\usepackage{amsthm}

\usepackage{import}
\usepackage{comment}
\usepackage{hyperref}
\usepackage{amsmath}
\usepackage{physics}
\usepackage{graphics}
\usepackage{caption}
\usepackage{subcaption}
\DeclareMathAlphabet{\pazocal}{OMS}{zplm}{m}{n}
\usepackage{collcell}
\usepackage{colortbl}
\usepackage{listings}
\usepackage{float}
\usepackage{xifthen}
\usepackage{xparse}

\usepackage{cleveref}
\Crefformat{figure}{Fig.\,#2#1#3}
\Crefmultiformat{figure}{Figs.\,#2#1#3}{ and~#2#1#3}{, #2#1#3}{ and~#2#1#3}
\crefrangeformat{figure}{Figs.\,#3#1#4 to~#5#2#6}
\Crefformat{section}{Sec.\,#2#1#3}
\Crefformat{equation}{Eq.\,(#2#1#3)}
\usepackage[acronym]{glossaries}

\usepackage{placeins}

\newacronym{FKS}{FKS}{Frixione-Kunst-Signer}
\newacronym{DGLAP}{DGLAP}{Dokshitzer-Gribov-Lipatov-Altarelli-Parisi}

\newacronym{LHC}{LHC}{Large Hadron Collider}
\newacronym{RHIC}{RHIC}{Relativistic Heavy Ion Collider}

\newacronym{PDF}{PDF}{parton distribution function}
\newacronym{FF}{FF}{fragmentation function}
\newacronym{PS}{PS}{parton shower}
\newacronym{POWHEG}{POWHEG}{Positive Weight Hardest Emission Generator}

\newacronym{IR}{IR}{infrared}
\newacronym{UV}{UV}{ultraviolet}
\newacronym{IRC}{IRC}{infrared collinear}

\newacronym{LO}{LO}{leading-order}
\newacronym{NLO}{NLO}{next-to-leading-order}
\newacronym{NLL}{NLL}{next-to-leading-logarithmic}
\newacronym{NNLO}{NNLO}{next-to-next-to-leading-order}
\newacronym{MiNLO}{MiNLO}{Multi-Scale Next-to-Leading Order}

\newacronym{QGP}{QGP}{quark-gluon plasma}
\newacronym{QCD}{QCD}{quantum chromodynamics}
\newacronym{QED}{QED}{quantum electrodynamics}

\newacronym{AJ}{AJ}{Albright-Jarlskog}

\NewDocumentCommand{\F}{O{} O{} m}{\ifthenelse{\isempty{#1}}
	{\ensuremath{F_{#3}^{#2}}}
	{\ensuremath{F_{#3,#1}^{#2}}}}

\usepackage{ulem} 

\usepackage{xcolor}
\usepackage{duckuments}

\title{Heavy-quark contributions to the polarized DIS structure functions at NLO in the ACOT scheme
}

\author[a]{E. Spezzano,}
\author[a]{T.~Je\v{z}o,}
\author[a]{M.~Klasen,}
\author[b]{I.~Schienbein}

\affiliation[a]{Institut  für  Theoretische  Physik,  Universität  Münster,  Wilhelm-Klemm-Straße 9, 48149 Münster, Germany}
\affiliation[b]{Laboratoire de Physique Subatomique et de Cosmologie, Université Grenoble-Alpes,CNRS/IN2P3, 53 Avenue des Martyrs, 38026 Grenoble, France}
\emailAdd{edoardo.spezzano@uni-muenster.de}
\emailAdd{tomas.jezo@uni-muenster.de}
\emailAdd{michael.klasen@uni-muenster.de}
\emailAdd{schien@lpsc.in2p3.fr}

\abstract{This study explores the heavy-quark contributions to polarized structure functions in deep-inelastic scattering at next-to-leading order. The structure functions \(g_1\), \(g_4\), \(g_5\), \(g_6\), and \(g_7\) are computed, while \(g_2\) and \(g_3\) are excluded due to the higher-twist suppression. The calculations are performed within the ACOT renormalization scheme, which ensures theoretical consistency across kinematic regions where heavy quarks transition from being dynamically produced to fully active degrees of freedom. By incorporating heavy-flavor contributions at next-to-leading-order, this work provides deeper insights into their role in polarized structure functions and the spin-dependent dynamics of QCD. Both analytical results and their numerical implementation are presented.}  

\preprint{MS-TP-26-01}


\begin{document}
\maketitle
\flushbottom 


\section{Introduction}
\label{sec:intro}
The spin structure of the nucleon remains one of the central open questions in hadronic physics and a key testing ground for quantum chromodynamics (QCD) across perturbative and non-perturbative scales~\cite{Filippone:2001ux, Leader_2010}. Polarized deep-inelastic scattering (DIS) has played a crucial role in this program by constraining the helicity-dependent parton distribution functions (PDFs) of quarks and gluons, and thereby quantifying the partonic contributions to the nucleon spin~\cite{Aidala_2013, de_Florian_2009}.

As we move into the era of high-precision nuclear physics, the forthcoming Electron-Ion Collider (EIC) in the United States~\cite{Accardi:2012qut, AbdulKhalek:2021gbh} and the proposed Electron-ion Collider in China (EicC)~\cite{Anderle_2021} are poised to deliver measurements of spin-dependent observables with unprecedented luminosity and kinematic reach. These facilities will catalyze a new generation of global QCD analyses~\cite{Nocera:2014gqa, Cocuzza:2022jye,Khanpour:2026erj}, making it imperative to achieve a commensurate level of theoretical accuracy. Central to this requirement is the rigorous treatment of heavy quark flavors—specifically charm and bottom—whose mass effects significantly influence the structure functions in the kinematic regimes targeted by these new colliders. 

Unpolarized heavy-quark structure functions are well established at next-to-leading order~\cite{laenenCompleteOaSCorrections1993} and next-to-next-to-leading order~\cite{Buza:1996ie}. By contrast, the polarized sector has only recently begun to receive comparable attention regarding detailed heavy-flavor dynamics~\cite{Hekhorn:2019nlf,Hekhorn:2024tqm}, following the pioneering next-to-leading order (NLO) studies~\cite{gluckModelsPolarizedParton2001}.

To systematically incorporate these effects, several general-mass variable-flavor-number schemes (GM-VFNS) have been developed—most notably the ACOT~\cite{Aivazis:1993kh, Aivazis:1993pi,Collins:1998rz}, FONLL~\cite{Forte:2010ta, ballCharmDeepinelasticScattering2015c}, and TR\mbox{~\cite{Thorne:1998ga,Thorne:2006qt}} frameworks—which ensure a theoretically consistent transition from the massive threshold region to the asymptotic limit where heavy quarks are treated as active partons. 

In this work, we build upon the current theoretical framework of polarized deep-inelastic scattering by presenting a generalized NLO treatment that incorporates full heavy-quark mass effects. While foundational benchmarks have often utilized the massless approximation~\cite{Vogelsang_1996,stratmannSpindependentNonsingletStructure1996} or focused on specific neutral-current (NC) analysis~\cite{Hekhorn:2019nlf}, our approach extends these results to a consistent massive description across all relevant NC and charged-current (CC) channels. At the order considered here, heavy-quark production is governed by boson-gluon fusion (GF) and quark scattering (QS) processes. We provide explicit analytic NLO results for the full massive coefficient functions contributing to the structure functions $g_1, g_4, g_5, g_6,$ and $g_7$. 
The transverse structure functions $g_2$ and $g_3$ are omitted from this analysis; their twist-2 components are determined by $g_1$ and $g_4$ via the Wandzura-Wilczek relations~\cite{Wandzura:1977qf}, while higher-twist contributions~\cite{Shuryak:1981pi} remain suppressed by powers of $M/Q$ in the Bjorken limit. 
The numerical evaluation of our results relies on a dedicated \texttt{Mathematica} code developed for this study. The implementation employs \texttt{ManeParse}~\cite{clarkManeParseMathematicaReader2017} to allow for an interface with the \texttt{NNPDFpol1.1} polarized PDF sets~\cite{Nocera:2014gqa}.

\section{Kinematics and formalism of polarized DIS}
\label{sec:kinematics}

In this work, we adopt standard kinematic conventions for polarized DIS involving heavy-quark production. We denote the four-momentum of the exchanged virtual boson as $q$, with the space-like momentum transfer $Q^2 = -q^2 > 0$. The initial- and final-state quarks, with masses $m_1$ and $m_2$, carry four-momenta $k_1$ and $k_2$, respectively. In the boson-gluon fusion process, the incoming gluon carries four-momentum $p$, while the incoming nucleon carries four-momentum $P$. 

The most general hadronic tensor consistent with Lorentz invariance is decomposed as~\cite{Hekhorn:2019nlf}
\begin{equation}
\begin{aligned}
 W_{\mu \mu^{\prime}}^{j}(S) &= -g_{\mu \mu^{\prime}} F_1^{j} + \frac{P_{\mu} P_{\mu^{\prime}}}{P \cdot q} F_2^{j} - i \varepsilon_{\mu \mu^{\prime} \alpha \beta} \frac{q^\alpha P^\beta}{2P \cdot q} F_3^{j} + \frac{q_\mu q_{\mu^{\prime}}}{Q^2} F_4^j + \frac{q_\mu P_{\mu^{\prime}} + q_{\mu^{\prime}} P_{\mu}}{2 P \cdot q} F_5^{j} \\
 &+ i \varepsilon_{\mu \mu^\prime \alpha \beta} \frac{q^\alpha}{P \cdot q} \left[ S^\beta g_1^j+ \left( S^\beta - \frac{S \cdot q}{P\cdot q} P^\beta \right) g_2^j \right] + \frac{1}{P \cdot q} \left[ \frac{1}{2} ( P_{\mu}^\perp S_{\mu^\prime}^\perp + S_\mu^\perp P_{ \mu^\prime}^\perp ) - \frac{S \cdot q}{P \cdot q} P_{\mu}^\perp P_{\mu^\prime}^\perp \right] g_3^j\\
 &+ \frac{S \cdot q}{P \cdot q} \left[ \frac{P_{\mu}^\perp P_{ \mu^{\prime}}^\perp}{P \cdot q} g_4^j+ \left( -g_{\mu \mu^{\prime}} + \frac{q_\mu q_{\mu^{\prime}}}{q^2} \right) g_5^j \right] + \frac{S \cdot q}{P \cdot q} \left[ \frac{q_\mu q_{\mu^{\prime}}}{q^2} g_6^j + \frac{q_\mu P_{\mu^{\prime}} + q_{\mu^{\prime}} P_{\mu}}{2 P \cdot q} g_7^j \right],
\end{aligned}
\label{eq:full_tensor}
\end{equation}
where $j \in \{\gamma, \gamma Z, Z, W\}$ labels the gauge boson exchange and $S^\mu$ is the spin four-vector of the nucleon. The transverse components are defined by $P_{\mu}^\perp = P_{\mu} - (P \cdot q / q^2) q_\mu$ and $S_\mu^\perp = S_\mu - (S \cdot q / q^2) q_\mu$. Here, $\varepsilon_{\mu \mu^\prime \alpha \beta}$ is the Levi-Civita tensor, $g_{\mu \mu^\prime}$ is the metric tensor, and the coefficients $F_i$ and $g_i$ represent the unpolarized and polarized structure functions, respectively.
Following established literature~\cite{Kretzer:1998ju, Blumlein:1996vs}, where applicable in subsequent sections, partonic quantities are distinguished from hadronic ones by the use of a hat symbol (e.g., $\hat{g}_i$ vs. $g_i$).

The identification of $g_2$ and $g_3$ as higher-twist functions is manifest in their coupling to the transverse projectors. These structures isolate sub-leading kinematic contributions that vanish in the collinear limit and are suppressed by powers of $1/Q$ relative to the leading-twist terms.

To isolate the spin-dependent dynamics in longitudinally polarized DIS, we define the antisymmetric component of the hadronic tensor as
\begin{equation}
 \Delta W_{\mu \nu}^j = \frac{1}{2} \left[ W_{\mu \nu}^j (-S) - W_{\mu \nu}^j(S) \right].
 \label{delta equation}
\end{equation}

While we also provide analytical expressions for the structure functions $g_6$ and $g_7$, our primary focus lies on $g_1$, $g_4$, and $g_5$, which directly enter the polarized cross section and are the main observables in our numerical analysis. In the limit of massless incoming leptons (e.g., electrons at the EIC), the contributions from $g_6$ and $g_7$ to the cross section are suppressed by factors of ($m_\ell^2/Q^2$), where $m_\ell$ denotes the incoming lepton mass. For electron beams, as relevant for the EIC, $(m_e^2/Q^2)$ is negligible in the kinematic region of interest, so the contributions from $g_6$ and $g_7$ to the cross section can safely be ignored. It is important to notice that these functions—in analogy with $F_4$ and $F_5$ in the unpolarized case~\cite{spezzanoHeavyquarkContributionsDIS2025}—are only non-negligible for massive leptons such as the $\tau$.

The polarized neutral-current and charged-current cross sections are given by~\cite{ParticleDataGroup:2022pth}
\begin{align}
&\frac{d^2 \Delta \sigma^{\rm NC}}{d x d y}=\frac{2 \pi y \alpha^2}{Q^4} \sum_{j=\gamma, \gamma Z, Z} \eta_j L_j^{\mu \nu} \Delta W_{\mu \nu}^j, \nonumber\\
&\frac{d^2 \Delta \sigma^{\rm CC}}{d x d y}=\frac{2 \pi y \alpha^2}{Q^4}  \eta_W L_W^{\mu \nu} \Delta W_{\mu \nu}^W,
\end{align}
where $\alpha$ is the fine structure constant,  the variables $x$ and $y$ are the standard Bjorken scaling variables $x = Q^2/(2P \cdot q)$ and $y = (P \cdot q)/(P \cdot k)$, and where the leptonic tensors are defined as
\begin{align}
L_{\mu \nu}^\gamma & =2\left(k_\mu k_\nu^{\prime}+k_\mu^{\prime} k_\nu-\left(k \cdot k^{\prime}-m_{\ell}^2\right) g_{\mu \nu}-i \lambda \varepsilon_{\mu \nu \alpha \beta} k^\alpha k^{\prime \beta}\right), \nonumber\\
L_{\mu \nu}^{\gamma Z} & =\left(g_V^e+e \lambda g_A^e\right) L_{\mu \nu}^\gamma, \quad L_{\mu \nu}^Z=\left(g_V^e+e \lambda g_A^e\right)^2 L_{\mu \nu}^\gamma,\nonumber \\
L_{\mu \nu}^W & =(1+e \lambda)^2 L_{\mu \nu}^\gamma.
\end{align}
Here, $k$ and $k^\prime$ are the momenta of the incoming and outgoing leptons, respectively, $\lambda$ denotes the lepton helicity, and $e = \pm 1$ represents the charge of the incoming lepton in units of the elementary charge. The channel-specific coefficients $\eta_j$ are defined as
\begin{equation}
\begin{aligned}
 \eta_\gamma &= 1, \quad & \eta_{\gamma Z} &= \left(\frac{G_F M_Z^2}{2 \sqrt{2} \pi \alpha}\right)\left(\frac{Q^2}{Q^2+M_Z^2}\right), \\
 \eta_Z &= \eta_{\gamma Z}^2, \quad & \eta_W &= \frac{1}{2}\left(\frac{G_F M_W^2}{4 \pi \alpha} \frac{Q^2}{Q^2+M_W^2}\right)^2.
 \label{eq:eta factors}
\end{aligned}
\end{equation}
The axial-vector ($g_A^e$) and vector ($g_V^e$) couplings for an incoming charged lepton are given by
\begin{equation}
g_{A}^{e} = -\frac{1}{2}, \quad g_{V}^{e} = -\frac{1}{2} + 2 \sin^2 \theta_W,
\label{eq:couplings}
\end{equation}
where $G_F$ is the Fermi constant, $M_Z$ and $M_W$ are the gauge boson masses, and $\theta_W$ is the Weinberg angle.

Considering the Bjorken limit~\cite{BjorkenLimit} (i.e $M^2\ll Q^2$), the polarized cross section simplifies to
\begin{equation}
\frac{d^2 \Delta \sigma^j}{d x d y} = \frac{2 \pi \alpha^2}{x y Q^2} \xi^j \left\{ -[1+(1-y)^2] g_4^j + y^2 g_L^j + 2 x [1-(1-y)^2] g_1^j \right\},
\label{eq:cross_section_final}
\end{equation}
where $\xi^{\text{NC}}=1$, $\xi^{\text{CC}}= 4 \eta_W$, and we define the longitudinal-like structure function $g_L^j = g_4^j - 2 x g_5^j$. Note that our normalization in~\cref{delta equation} introduces a factor of 2 difference compared to the PDG convention to ensure that experimental single-spin asymmetries are bounded by unity~\cite{Aschenauer_2013}. 
Finally, the total hadronic structure functions are given by~\cite{borsaFullSetPolarized2022} 
\begin{align}
 &g_{\{1,4,5\}}^{\text{CC}} = g_{\{1,4,5\}}^{W^\pm}, \nonumber\\
 &g_1^{\text{NC}} = g_1^\gamma - g_V^e \eta_{\gamma Z} g_1^{\gamma Z} + \left( g_V^{e \, 2} + g_A^{e \, 2} \right) \eta_Z g_1^Z, \nonumber\\
 &g_{\beta}^{\text{NC}} = g_A^e \eta_{\gamma Z} g_\beta^{\gamma Z} - 2 g_V^e g_A^e \eta_Z g_\beta^Z,\quad \beta= 4, \, 5.
\end{align}

\section{Quark scattering}

To compute the contribution of the quark scattering process
\begin{equation}
B^*(q) + Q_1(k_1) \to Q_2(k_2) + g(p_2),
\end{equation}
where \(B^*\) denotes a virtual electroweak boson (\(\gamma, Z, W^\pm\)), \(g\) the emitted gluon with momentum \(p_2\), and \(k_1\) (\(k_2\)) the momentum of the incoming (outgoing) quark with mass \(m_1\) (\(m_2\)), the contribution to the structure functions receives two components: the real emission and the virtual corrections.

\begin{figure}[H]
    \centering
    \includegraphics[scale=0.8]{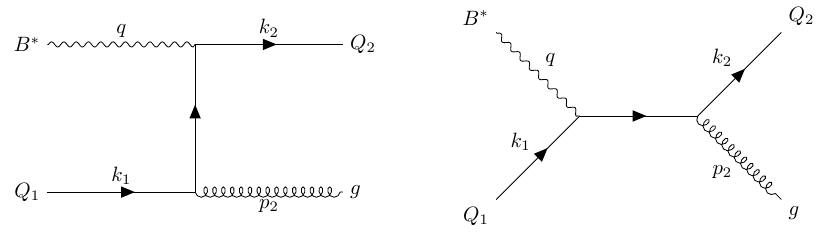}
\caption{Real-emission diagrams for the quark-scattering channel $B^*(q) + Q_1(k_1) \to Q_2(k_2) + g(p_2)$ at $\mathcal{O}(\alpha_s)$.
Gluon radiation off the incoming and outgoing heavy-quark lines generates the polarized partonic structure functions $\hat g_i^{\rm QS}$ with full dependence on the heavy-quark masses $m_1$ and $m_2$.}
\label{fig:Real Emission Diagram}
\end{figure}

\begin{figure}[H]
    \centering
    \includegraphics[scale=0.65]{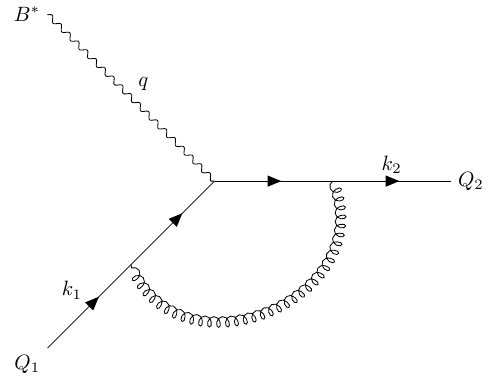}
    \caption{
    One-loop vertex correction to the quark-boson vertex in the quark-scattering process $B^*(q) + Q_1(k_1) \to Q_2(k_2)$.
    The calculation is performed in the on-shell renormalization scheme, where self-energy insertions on the external heavy-quark legs are canceled by the corresponding mass and wave-function counterterms. Together with the real-emission diagrams in~\cref{fig:Real Emission Diagram}, this contribution renders the NLO polarized
structure functions infrared finite.
}
\label{fig:Vertex Correction}
\end{figure}

For real emission, the relevant Feynman diagrams are shown in~\cref{fig:Real Emission Diagram}. These diagrams represent the processes involving the emission of an additional gluon in the final state and thus contribute to the scattering process at the next-to-leading order. On the other hand, the virtual contribution is represented by the vertex correction, as illustrated in~\cref{fig:Vertex Correction}. In the on-shell renormalization scheme, self-energy insertions on external fermions legs are exactly canceled by the corresponding mass and wave-function renormalization counterterms and therefore do not appear explicitly in the diagram \cite{peskinIntroductionQuantumField1995}. These two contributions must be summed at the structure function/cross section level to cancel the 
poles, obtaining a consistent result~\cite{Kinoshita:1962ur,1964PhRv..133.1549L}.

\subsection{Real contribution}

To evaluate the real radiation contributions, we begin by factorizing the scattering amplitude as\begin{equation}
\mathcal{M}^{\rm QS} = \varepsilon^{\mu}(q) \mathcal{M}_{\mu}^{\rm QS} u(k_1,h) ,
\end{equation}
where $\varepsilon^{\mu}(q)$ is the polarization vector of the incoming gauge boson $B^*$ with four-momentum $q$. The term $u(k_1, h)$ is the Dirac spinor representing the initial-state quark with four-momentum $k_1$ and helicity $h = \pm 1$. The term $\mathcal{M}_{\mu}^{\rm QS}$ denotes the reduced amplitude once the external virtual boson polarization vector $\varepsilon^{\mu}(q)$ and the quark spinor are factored out. The partonic tensor for the hard-scattering process (prior to phase space integration) is then given by
\begin{equation}
\tilde{W}_{\mu \nu}^{\rm QS} (h) =  \mathcal{M}_{\mu}^{\rm QS}(V, A) \left[ u(k_1, h) \bar{u}(k_1, h) \right] \left(\mathcal{M}_{\nu}^{\rm QS}(V^\prime, A^\prime) \right)^* ,
\end{equation}
where $V, A$ and $V', A'$ represent the vector and axial-vector couplings associated with the first and second current insertions, respectively.

As is well known, averaging over the helicity $h$ of the incoming quark leads to the unpolarized case. Therefore, in our analysis, it is crucial to explicitly maintain the helicity dependence. Specifically, considering an incoming quark, we have\footnote{The equation presented in our paper differs slightly from the one given in the referenced work due to a different choice of convention. In particular, the two expressions can be matched by noting that \( h = 2\lambda \) and \( s = h \hat{S} / m_1 \).} (see eq.~(1.51) in~\cite{Haber:1994pe})
\begin{equation}
    u(k_1, h) \bar{u}(k_1, h)= \frac{1}{2}\left( 1+ h \gamma_5 \frac{\hat{\slashed{S}}}{m_1}\right) ( \not{k}_1+m_1)\, ,
    \label{eq:uubar}
\end{equation}
where $\hat{S}^{\mu}$ denotes the spin four-vector of the fermion. The equation given above reduces to eq.~(2.13) in~\cite{bojak2000nloqcdcorrectionspolarized} only in the ultrarelativistic limit. Retaining the full mass dependence in~\cref{eq:uubar} is not merely a matter of precision; it is essential for the consistent treatment of infrared behaviors. Specifically, this ensures the proper cancelation of poles as prescribed by the KLN theorem~\cite{Kinoshita:1962ur}, which guarantees that the total cross section remains infrared-safe when all degenerate states are summed.

It is important to emphasize that in our computation $\gamma_5$ is treated strictly in $D=4$. Since we perform a fully massive calculation, no $D$-dimensional definition of $\gamma_5$ is required, and the projectors are constructed explicitly in four dimensions (see \cref{Projectors appendix} for details). The heavy-quark mass provides a physical cutoff for collinear divergences, which justifies the use of four-dimensional projectors. In contrast, the phase-space integration is performed in $D$ dimensions\mbox{~\cite{Kretzer:1998ju, spezzanoHeavyquarkContributionsDIS2025}} to regulate soft singularities and ensure the correct cancellation of infrared poles.

Having introduced the helicity projector, we now specify the definition of the spin four-vector. In the standard light-cone formalism~\cite{Mulders_1996}, it is defined as
\begin{equation}
\hat{S}^\mu \propto  \left( \frac{k_1^\mu}{m_1} - \frac{m_1}{k_1 \cdot n} n^\mu \right).
\end{equation}
For the purpose of the partonic tensor computation, it is useful to project $\hat{S}^\mu$ into the $\{k_1, q\}$ basis. Utilizing the kinematic identities $\hat{S} \cdot k_1 = 0$ and $\hat{S}^2 = -m_1^2$, and using the factorization theorem~\cite{Collins:1998rz}, the spin four-vector is then decomposed as
\begin{equation}
   \hat{S}^\mu =  \frac{Q^2}{\hat{x} \Delta'} \left( k_1^\mu - 2 \hat{x} \frac{m_1^2}{Q^2} q^\mu \right),
\end{equation}
where the variables $\Delta^\prime$ and $\hat{x}$ will be defined later in~\cref{deltaprime} and~\cref{xhat} as functions of partonic and kinematic variables.
This decomposition is particularly useful for expressing scalar products involving the spin four-vector in terms of Mandelstam variables.

Finally, taking the difference as explained in~\cref{sec:kinematics}, we define the following quantity:
\begin{equation}
    \Delta \hat{W}_{\mu \nu}^{\rm QS}= \frac{1}{2} \left( \tilde{W}_{\mu \nu}^{\rm QS}(-1)- \tilde{W}_{\mu \nu}^{\rm QS}(1) \right)\, .
    \label{eq:DeltaWhatQS}
\end{equation}
Then, by applying the projectors specified in~\cref{Projectors appendix} at the partonic level and following the same conventions as in~\cite{Kretzer:1998ju,spezzanoHeavyquarkContributionsDIS2025},
the polarized partonic structure functions are obtained as:
\begin{equation}
\hat{g}_{i}^{\rm QS} \equiv \left(C_F \frac{g_{s}^2}{2}\right)^{-1} \hat{P}^{\mu \nu}_{i, \rm QS} \overline{\Delta \hat{W}_{\mu \nu}^{\rm QS}} , \quad   i \in \{1, 4, 5, 6, 7\} ,
\end{equation}
where $C_F = 4/3$, $g_s$ is the strong coupling constant, and 
\begin{equation}
    \overline{\Delta \hat{W}_{\mu \nu}^{\rm QS}} = \frac{1}{3} \sum_{\rm color} \Delta \hat{W}_{\mu \nu}^{\rm QS}.
    \label{average tensor boson gluon}
\end{equation}
In this expression, the factor of $1/3$ originates from averaging over the color states of the incoming quark.

In order to write down the partonic structure functions in a compact manner, it is useful to define the following
quantities:
\begin{equation}
  q_{\pm} = V V^\prime \pm A A^\prime, \ R_+= \frac{AV^\prime+ A^\prime V}{2}, \  \Sigma_{\pm \pm} = Q^2 \pm m_2^2 \pm m_1^2, \ \hat{t}_1=(k_2 - q)^2 -m_{1}^2  , \ \hat{s}_1= (k_1 + q)^2-m_{2}^2
  \label{s1 expression}
\end{equation}
and
\begin{equation}
  \Delta \equiv \Delta(m_1^2, m_2^2, -Q^2), \quad \Delta^{\prime} \equiv \Delta( m_1^2,\hat{s}_1+m_2^2, -Q^2)
  \label{deltaprime}
\end{equation}
with the triangle function
\begin{equation}
    \Delta(a, b, c) = \sqrt{a^2 + b^2 + c^2 - 2ab - 2ac - 2bc}.
\end{equation}
Moreover, the variable $\hat{x}$ can be expressed in terms of $\hat{s}_1$ using the following equation:
\begin{equation}
    \hat{s}_1 = m_1^2-m_{2}^2+Q^2 \left( \frac{1}{\hat{x}} - 1 \right).
    \label{xhat}
\end{equation}
Using these definitions, the unintegrated partonic structure functions are given by
\begin{equation*}
\begin{aligned}
    \hat{g}_{1}^{\rm QS} = &  -\frac{4}{ \Delta^{\prime \, 2}} \Bigg\{ 2 \Delta^2 \Big[q_+ \Sigma_{++} \hspace{-0.2em} - \hspace{-0.2em} q_- 2 m_1 m_2 \Big] \left( \frac{m_2^2}{\hat{s}_1^2} + \frac{m_1^2}{\hat{t}_1^2} + \frac{\Sigma_{++}}{\hat{s}_1 \hat{t}_1} \right) \hspace{-0.2em} +\hspace{-0.2em}  2 \left( 2 \Delta^2 - \Sigma_{-+} \Sigma_{++} \right) \left[ q_+ \Sigma_{++} - q_- 2 m_1 m_2 \right] \frac{1}{\hat{s}_1}\\
    &+ \hspace{-0.3em} \frac{2 m_1^2 \hat{s}_1 \hspace{-0.2em} \left[ q_+ \hspace{-0.2em} \left( \Delta^{\prime \, 2} \hspace{-0.3em} + \hspace{-0.3em} \Delta^2\hspace{-0.3em} - \hspace{-0.3em} 2 m_1^2 (2 \Sigma_{-+} \hspace{-0.3em} - \hspace{-0.3em} \hat{s}_1) \right)\hspace{-0.2em} - \hspace{-0.2em} 2 m_1 m_2 q_- \hspace{-0.2em} \left( 2 \Sigma_{+-} \hspace{-0.2em} + \hspace{-0.2em} \hat{s}_1 \right) \right]}{\hat{t}_1^2} \hspace{-0.2em} +\hspace{-0.2em}  \left[ q_+ \hspace{-0.2em} \left( \Sigma_{++} \Sigma_{+-} \hspace{-0.2em} -\hspace{-0.2em}  4 m_1^2 m_2^2 \right) \hspace{-0.2em} + \hspace{-0.2em} 4 q_- m_1 m_2^3 \right] \frac{\hat{t}_1}{\hat{s}_1}\\
    &+ \left[ q_+ \left( \hat{s}_1 \Delta^{\prime \, 2} + 2 (\Sigma_{++} + m_1^2) \hat{s}_1^2 + (-\Delta^2 + 7 \Sigma_{++} \Sigma_{+-} + 4 m_1^2 m_2^2) \hat{s}_1 + 2 \Sigma_{++} \left( 3 \Delta^2 - 4 m_1^2 \Sigma_{-+} \right) \right) \right] \\
    &- q_- \left[ 4 m_1 m_2 \left( \Delta^{\prime \, 2} + \left( m_2^2 + Q^2 \right) \hat{s}_1 + 2 \Sigma_{+-} \Sigma_{++} \right) \right] \frac{1}{\hat{t}_1}- q_- \left[ 4 m_1 m_2 \left( -3 m_1^2 + 2 m_2^2 + Q^2 - \hat{t}_1 \right) \right] \\
    &+ q_+ \left[ 2 \left( (-2 m_1^2 + m_2^2) \hat{s}_1 - (2 m_1^2 + m_2^2) \hat{t}_1 + \Sigma_{++} \left( 2 \Sigma_{+-} - \Sigma_{-+} \right) - 2 m_1^2 m_2^2 \right) - \hat{s}_1 \hat{t}_1 \right]  \Bigg\},\\
\end{aligned}
\label{g1 paper}
\end{equation*}
\begin{equation*}
\begin{aligned}
    \hat{g}_4^{\rm QS} = &   -\frac{16 R_+ Q^4}{\hat{x}^2 \Delta^{\prime \, 6}}   \Bigg\{ 
     \hspace{-0.1em} 2 \Sigma_{+-} \Delta^4  \hspace{-0.2em} \left(  \hspace{-0.2em} \frac{m_2^2}{\hat{s}_1^2}  \hspace{-0.1em} +  \hspace{-0.1em}\frac{m_1^2}{\hat{t}_1^2}  \hspace{-0.1em} +  \hspace{-0.1em}\frac{\Sigma_{++}}{\hat{s}_1 \hat{t}_1}  \hspace{-0.2em} \right) \hspace{-0.2em} -   \hspace{-0.2em} \hat{t}_1 \Delta^2 \frac{\Sigma_{-+}( \Sigma_{++} -3 Q^2)+Q^2 \Sigma_{++}}{\hat{s}_1}  \hspace{-0.2em}  -  \hspace{-0.2em}16 \Sigma_{+-} Q^2 m_1^2 m_2^2\\
    &+\hspace{-0.2em} 2 m_1^2 \hat{s}_1 \frac{  \Sigma_{+-} \hat{s}_1^3  \hspace{-0.2em}+  \hspace{-0.2em} 3 (\Delta^2 \hspace{-0.2em} -  \hspace{-0.2em} 2 m_1^2 Q^2) \Delta^{\prime \, 2}  \hspace{-0.2em} - \hspace{-0.2em}  6 m_1^2 Q^2 \Delta^2  \hspace{-0.2em} -  \hspace{-0.2em} 2 \Sigma_{+-} (\Delta^2  \hspace{-0.2em}-  \hspace{-0.2em} 2 m_1^2 Q^2) \hat{s}_1 }{\hat{t}_1^2} +\frac{\Delta^2 \Sigma_{+-} (5 \Delta^2 - 3 \Sigma_{++} \Sigma_{-+})}{\hat{s}_1}  \\
    &+ \frac{\Sigma_{+-} \hat{s}_1^4 + ((5 \Sigma_{+-} + 4 m_1^2) \Sigma_{++} - 8 m_1^2 m_2^2) \hat{s}_1^3 + ((11 \Sigma_{++} - 6 m_1^2) \Delta^2 - 2 m_1^2 Q^2 (17 \Sigma_{+-} + 14 m_1^2)) \hat{s}_1^2}{\hat{t}_1} \\
    &+  \hspace{-0.3em} \hat{s}_1 \frac{  (13 \Sigma_{+-}  \hspace{-0.2em} +  \hspace{-0.2em} 8 m_1^2) \Sigma_{++} \Delta^2 \hspace{-0.2em} - \hspace{-0.2em} 16 m_1^2 m_2^2 (m_2^2 \hspace{-0.2em}  - \hspace{-0.2em}  m_1^2)^2  \hspace{-0.2em} - \hspace{-0.2em}  2 m_1^2 Q^2 \left( 21 \Delta^2 \hspace{-0.2em}  - \hspace{-0.2em}  8 (m_2^2 Q^2 \hspace{-0.2em} - \hspace{-0.2em} \Sigma_{+-} \Sigma_{-+} \hspace{-0.2em}  + \hspace{-0.2em}  2 \Sigma_{-+} \Sigma_{++} \hspace{-0.2em} - \hspace{-0.2em}  \Sigma_{++}^2) \right)}{\hat{t}_1} \\
    &+ \hat{s}_1^2 \left( (5 \Sigma_{++} \hspace{-0.2em}  + \hspace{-0.2em}  3 \Sigma_{+-}) \Sigma_{++} - 20 m_1^2 m_2^2 - 12 \Sigma_{+-} Q^2 + \Sigma_{+-} \hat{t}_1 \right) + \frac{4 \left( (2 \Sigma_{+-} + 3 m_1^2) \Delta^2 - 6 m_1^2 Q^2 \Sigma_{++} \right) \Delta^2}{\hat{t}_1}\\
    &+ \hat{s}_1 \left( (2 \Delta^2 + (2 Q^2 + \Sigma_{+-}) \Sigma_{++} - 8 \Sigma_{+-} Q^2) \hat{t}_1 - 4 m_1^2 Q^2 (9 m_2^2 - 4 m_1^2) + (14 m_2^2 - 6 Q^2) \Delta^2 \right) \hspace{-0.2em}+ \hspace{-0.2em} 2 (m_2^2\hspace{-0.2em} -\hspace{-0.2em} m_1^2) \hat{s}_1^3\\
    &+ \hspace{-0.2em} \hat{t}_1 \hspace{-0.2em} \left(  \hspace{-0.1em} (2 \Sigma_{++} \hspace{-0.2em} - \hspace{-0.2em} 12 Q^2 \hspace{-0.2em} + \hspace{-0.2em} \Sigma_{+-}) \Delta^2 \hspace{-0.2em} - \hspace{-0.2em} Q^2 \Sigma_{++}^2 \hspace{-0.2em} + \hspace{-0.2em} Q^2 (6 Q^2 \hspace{-0.2em} + \hspace{-0.2em} \Sigma_{+-}) \Sigma_{++} \right) \hspace{-0.2em}+ \hspace{-0.2em} 2 \hspace{-0.2em} \left( 4 \Delta^4 \hspace{-0.2em} - \hspace{-0.2em} (3 \Sigma_{++} (\Sigma_{-+} \hspace{-0.2em} + \hspace{-0.2em} 2 Q^2)  \hspace{-0.2em}- \hspace{-0.2em} 5 Q^2 \Sigma_{+-})  \hspace{-0.1em}\right) \hspace{-0.2em} \Delta^2 \hspace{-0.2em} \Bigg\},\\
\end{aligned}
\label{g4paper}
\end{equation*}
\begin{equation*}
\begin{aligned}
    \hat{g}_5^{\rm QS} = &  - \frac{8 R_+ Q^2}{ \hat{x} \Delta^{\prime \,4}} \Bigg\{ 
    2 \Delta^4 \left( \frac{m_2^2}{\hat{s}_1^2} + \frac{m_1^2}{\hat{t}_1^2} + \frac{\Sigma_{++}}{\hat{s}_1 \hat{t}_1} \right) + \frac{\hat{s}_1^4 + (6 m_1^2 + 5 \Sigma_{+-}) \hat{s}_1^3 + (24 m_1^4 - 12 m_1^2 m_2^2 + 11 \Sigma_{+-} \Sigma_{++}) \hat{s}_1^2}{\hat{t}_1}\\
    &+ \frac{2 m_1^2 \hat{s}_1 \left( \hat{s}_1 (\Sigma_{+-}^2 - 4 m_1^2 (m_2^2 - m_1^2)) + \Delta^2 (3 \Sigma_{+-} - \Sigma_{++}) + (\hat{s}_1 + \Sigma_{++}) \Delta^{\prime \, 2} \right)}{\hat{t}_1^2}+ \frac{\Delta^2 (5 \Delta^2 - 3 \Sigma_{++} \Sigma_{-+})}{\hat{s}_1}  \\
    &+ \frac{\left((12 m_1^2 + 13 \Sigma_{+-}) \Delta^2 - 16 m_1^2 Q^2 \Sigma_{++}\right) \hat{s}_1 + 4 \left((2 \Sigma_{+-} + m_1^2) \Sigma_{++} \Delta^2 - 2 m_1^2 m_2^2 \Delta^2 \right)}{\hat{t}_1} + \frac{\Delta^2 \Sigma_{-+} \hat{t}_1}{\hat{s}_1}\\
    &+ 2 \left((-3 Q^2 + \Sigma_{++} + 4 \Sigma_{+-}) \Delta^2 - 8 m_1^2 m_2^2 Q^2 \right)+ \hat{t}_1 \Big(\hat{s}_1 (\Sigma_{++} + 4 m_1^2 + \hat{s}_1) + (\Sigma_{-+} - 4 m_1^2) \Sigma_{++} + 12 m_1^2 m_2^2 \Big)\\
    &+ \hat{s}_1 \left(10 m_1^2 \hat{s}_1 + \left(5 \Sigma_{++}^2 - (6 Q^2 + \Sigma_{+-}) \Sigma_{++} + 4 Q^2 \Sigma_{+-} - 8 m_1^2 m_2^2 \right)\right)\Bigg\}, \\
\end{aligned}
\label{g5paper}
\end{equation*}
\begin{equation*}
\begin{aligned}
    \hat{g}_6^{\rm QS} = &  -\frac{8 R_+}{\hat{x} \Delta^{\prime \, 4}} \Big\{ 
    2 \Delta^2 \Big[ (m_2^2-m_1^2) \Delta^2+4 m_1^2 Q^2 \Sigma_{-+} \Big] \Big( \frac{m_2^2}{\hat{s}_1^2} + \frac{m_1^2}{\hat{t}_1^2} + \frac{\Sigma_{++}}{\hat{s}_1 \hat{t}_1} \Big) + \frac{\big[(5 m_2^2 + m_1^2) \Sigma_{+-} + 2 m_1^2 Q^2\big] \hat{s}_1^3}{\hat{t}_1}\\
    &+ \frac{\big[(m_1^2 \hspace{-0.2em} -\hspace{-0.2em}  m_2^2)(-\Sigma_{++} \Delta^2 \hspace{-0.2em}+ \hspace{-0.2em} 4 Q^2 \Delta^2) \hspace{-0.2em} +\hspace{-0.2em}  8 m_1^2 m_2^2 Q^2 \Sigma_{-+}\big] \hat{t}_1}{\hat{s}_1} \hspace{-0.2em} +\hspace{-0.2em}  \frac{\big[(5 m_1^2 \hspace{-0.2em} +\hspace{-0.2em}  11 m_2^2) \Sigma_{+-}^2 \hspace{-0.2em} + \hspace{-0.2em} 4 m_1^2 Q^2 (7 m_1^2\hspace{-0.2em} -\hspace{-0.2em} m_2^2\hspace{-0.2em} +\hspace{-0.2em} 4 Q^2)\big] \hat{s}_1^2}{\hat{t}_1} \\
    &+ \frac{2 (4 m_2^2 - m_1^2) \Delta^4 + Q^2 \big[2 (4 m_1^2 - 3 m_2^2) \Sigma_{++} \Delta^2 + 4 m_1^2 m_2^2 (-9 \Delta^2 + 8 Q^2 \Sigma_{++})\big]}{\hat{s}_1} + \frac{(m_2^2 - m_1^2) \hat{s}_1^4}{\hat{t}_1} \\
    &+ \frac{\Big[(13 m_2^2 - 5 m_1^2) \Sigma_{++} \Delta^2 + 8 m_1^2 \big(-2 m_2^2 (m_2^2 - m_1^2)^2 + Q^2 \Sigma_{++} (2 m_1^2 - 9 m_2^2 + 5 Q^2)+2 m_2^2 Q^2(3m_1^2+Q^2)\big)}{\hat{t}_1} \\
    &+ 2 (m_2^2 - m_1^2) \hat{s}_1^3 + \hat{s}_1^2 \Big[(5 \Sigma_{++} + 3 \Sigma_{+-}) \Sigma_{++} - 20 m_1^2 m_2^2 - 10 m_1^2 Q^2 - 12 \Sigma_{+-} Q^2 + 4 Q^4 + (\Sigma_{+-} - Q^2) t_1 \Big] \\
    &+ \hat{s}_1 \Big[(2 \Delta^2 + (2 Q^2 + \Sigma_{+-}) \Sigma_{++} - Q^2 (3 \Sigma_{++} + 6 \Sigma_{+-} - 4 Q^2)) \hat{t}_1 + \big((7 \Sigma_{++} + 7 \Sigma_{+-} + 3 Q^2) \Delta^2 \\
    & - Q^2 (8 m_1^2 Q^2  \hspace{-0.2em} + \hspace{-0.2em} 32 m_1^4 \hspace{-0.2em} + \Sigma_{++} (-5 \Sigma_{-+} \hspace{-0.2em} + \hspace{-0.2em} 22 \Sigma_{+-}))\big)\Big] + \hat{t}_1 \Big[(2 \Sigma_{++} \hspace{-0.2em} + \hspace{-0.2em} \Sigma_{+-} \hspace{-0.2em}- \hspace{-0.2em} 5 Q^2) \Delta^2 \hspace{-0.2em} - \hspace{-0.2em} 2 Q^2 \big(\Sigma_{+-} (-2 \Sigma_{-+}\hspace{-0.2em}  + \hspace{-0.2em} \Sigma_{++})\\
    & + 4 m_1^2 m_2^2 - 8 m_1^2 Q^2\big)\Big] + \frac{\Big[(8 m_2^2 + 4 m_1^2) \Delta^4 + 4 m_1^2 Q^2 \big((-3 m_1^2 +5 Q^2 - 13 m_2^2) \Delta^2 + 32 m_1^2 m_2^2 Q^2\big)\Big]}{\hat{t}_1} \\
    &+ 2 \Big[4 \Delta^4 - (9 m_1^2 Q^2 + m_2^2 Q^2 + 2 Q^4 + 3 \Sigma_{-+} \Sigma_{++}) \Delta^2 + Q^2 \Sigma_{++} (\Sigma_{++} (3 \Sigma_{-+} - 2 \Sigma_{+-}) + 16 m_1^2 Q^2)\Big] \\
    &+ \frac{2 m_1^2 \hat{s}_1}{\hat{t}_1^2} \Bigg[\hspace{-0.2em} (m_2^2 - m_1^2) \hat{s}_1^3 \hspace{-0.2em}+ \hspace{-0.2em}\big(3 (\Delta^2  \hspace{-0.2em} -  \hspace{-0.2em} Q^2 \Sigma_{++}) \hspace{-0.2em}- \hspace{-0.2em}2 m_1^2 Q^2\big) s_1^2\hspace{-0.2em}+\hspace{-0.2em} \big(\Delta^2 (3 \Delta^2 \hspace{-0.2em}- \hspace{-0.2em} 8 Q^2 \Sigma_{++}  \hspace{-0.2em} +  \hspace{-0.2em} 5 Q^4) + 8 Q^4 (m_2^2 + m_1^2) \Sigma_{++}\big)  \\
    &\quad + 2 \big((2 \Sigma_{+-} + Q^2) \Delta^2 - 3 Q^2 \Sigma_{++} \Sigma_{+-}\big) \hat{s}_1 + Q^2 (m_2^2 - m_1^2) \big(5 \Delta^2 - 8 Q^2 \Sigma_{++}\big) \Bigg] \Bigg\}, 
   \end{aligned}
\label{g6paper}
\end{equation*}
\begin{equation*}
\begin{aligned}
    \hat{g}_7^{\rm QS} = & \frac{16 R_+ Q^2 }{\hat{x}^2 \Delta^{\prime \, 4}} \Bigg\{ 
    2 \Delta^2 (\Delta^2 - Q^2 \Sigma_{++}) \Bigg( \frac{m_2^2}{\hat{s}_1^2} + \frac{m_1^2}{\hat{t}_1^2} + \frac{\Sigma_{++}}{\hat{s}_1 \hat{t}_1} \Bigg) + \frac{2 \big[(\Sigma_{+-} - 4 Q^2) \Sigma_{++} \Delta^2 + 2 \Delta^4 + \Sigma_{-+} \Sigma_{++}^2 Q^2\big]}{\hat{s}_1} \\
    &+ \hspace{-0.2em} \frac{2 m_1^2 \hat{s}_1}{\hat{t}_1^2} \Bigg[ \hspace{-0.2em} (m_2^2 - m_1^2) \hat{s}_1^2 + 2 (\Delta^2 \hspace{-0.2em}-  \hspace{-0.2em} Q^2 (\Sigma_{++} \hspace{-0.2em} +  \hspace{-0.2em} m_1^2)) \hat{s}_1 + 2 \Sigma_{+-} (\Delta^2 \hspace{-0.2em} - \hspace{-0.2em}  Q^2 \Sigma_{++}) \Bigg] \hspace{-0.2em} + \hspace{-0.2em} \hat{t}_1 \frac{(m_1^2 + m_2^2) \Delta^2 - 2 m_2^2 Q^2 \Sigma_{++}}{\hat{s}_1}\\
    &+ \hat{s}_1 \frac{ \hspace{-0.1em}(m_2^2 - m_1^2) \Delta^{\prime \, 2} + (\Sigma_{++}^2 - 5 m_1^4 + m_2^4 - Q^4) \hat{s}_1 + 2 \big[(2 m_1^2 + 3 m_2^2) \Delta^2 + m_1^2 Q^2 (\Sigma_{-+} - 14 m_2^2)\big] \hspace{-0.2em}}{\hat{t}_1} \\
    &+ \hspace{-0.2em} \frac{\big[2 (3 m_2^2 - m_1^2) \Sigma_{++} \Delta^2 + 8 m_1^2 ((Q^2 - m_2^2) \Delta^2 - 2 m_2^2 Q^2 \Sigma_{+-})\big]}{\hat{t}_1}  \hspace{-0.2em} + \hspace{-0.1em}  \hat{t}_1 \Big[ \Delta^2   \hspace{-0.2em} -  \hspace{-0.2em} Q^4  \hspace{-0.1em} +  \hspace{-0.1 em}(m_2^2 - m_1^2) (\Sigma_{++} \hspace{-0.1em} - \hspace{-0.1em} 3 Q^2+ \hat{s}_1)\Big] \\
    &+ \hspace{-0.2em} \hat{s}_1 \Big[2 (m_2^2 \hspace{-0.2em} - \hspace{-0.2em} m_1^2) \hat{s}_1   \hspace{-0.2em} +  \hspace{-0.2em} 3 \Delta^2 \hspace{-0.3em}  - \hspace{-0.2em} (-3 \Sigma_{+-} \hspace{-0.2em} + \hspace{-0.2em} 2 Q^2) \Sigma_{++} - 2  Q^2 (3 m_2^2 + 2 Q^2)\Big] \hspace{-0.2em} + \hspace{-0.2em} 2 (m_2^2  \hspace{-0.2em} +  \hspace{-0.2em} Q^2) \big(4 \Delta^2 \hspace{-0.2em} - \hspace{-0.2em} Q^2 (5 \Sigma_{++} \hspace{-0.2em} - \hspace{-0.2em} \Sigma_{-+})\big)  \hspace{-0.2em} \Bigg\}\, .
\end{aligned}
\label{g7paper}
\end{equation*}

\subsection{Wilson coefficients and NLO contribution}

The Wilson coefficients are defined as
\begin{equation}
\Delta \hat{H}_{i}^{\rm QS, \, LO} \equiv \left(\frac{g_{s}^2}{2}\right)^{-1}  P_{i}^{\mu \nu} \overline{\Delta \hat{W}_{\mu \nu}^{\rm QS}} \Big|_{k_{1}^{+}= \xi P^{+}} , \quad i=1,\cdots,7\ ,
\label{mixing equation}
\end{equation}
where the hadron level projection operators $P_{i}^{\mu \nu}$ 
can be found in the~\cref{Projectors appendix}.
To facilitate the discussion, we introduce the partonic scaling variable $\xi^\prime= \chi/\xi$, where
\begin{equation}
    \chi= \frac{x}{2 Q^2} ( \Sigma_{+-}+\Delta)\, .
\end{equation}
Using the approximation $Q^2 \gg M^2$, where $M$ is the mass of the nucleon, and switching to the phase-space integrated 
version of~\cref{mixing equation}, we obtain
\begin{align}
\Delta \hat{\mathcal{H}}_{1}^{\rm QS, \, NLO} &= \mathfrak{g}_{1}^{\rm QS, \, NLO}, \nonumber
\\
\Delta\hat{\mathcal{H}}_{4}^{\rm QS, \, NLO} &=  \frac{\chi}{\xi^\prime} (1+\mathcal{R})  \left(\frac{1+\mathcal{R}}{1-\mathcal{R}} \right)^2 \mathfrak{g}_{4}^{\rm QS, \, NLO},
\nonumber\\
\Delta \hat{\mathcal{H}}_{5}^{\rm QS, \, NLO} &=  \frac{1+\mathcal{R}}{1-\mathcal{R}} \mathfrak{g}_{5}^{\rm QS, \, NLO},
\nonumber\\
\Delta \hat{\mathcal{H}}_{6}^{\rm QS, \, NLO} &= \frac{1+\mathcal{R}}{1-\mathcal{R}} \left( \mathfrak{g}_{6}^{\rm QS, \, NLO} - \frac{2 \mathcal{R}}{1-\mathcal{R}} \mathfrak{g}_{7}^{\rm QS, \, NLO}\right),
\nonumber\\
\Delta \hat{\mathcal{H}}_{7}^{\rm QS, \, NLO} &= \left(\frac{1+\mathcal{R}}{1-\mathcal{R}}\right)^2 \mathfrak{g}_{7}^{\rm QS, \, NLO},
\label{mixing equations}
\end{align}
where 
\begin{equation}
\mathfrak{g}_i^{\rm QS, \, NLO} =  C_F \left( N_i (S_i + V_i) \delta(1 - \xi') + \frac{1}{8} \frac{1 - \xi'}{(1 - \xi')_+} \frac{\hat{s}_1}{\hat{s}_1 + m_2^2}  \int_{0}^{1} \hat{g}_{i}^{\rm QS}( \hat{s}_1, \hat{t}_1(y)) \, dy \right)
\label{Inclusive QS}
\end{equation}
with the coefficients $N_i, \, S_i , \, V_i$ reported in~\cref{Appendix Virtual} and 
\begin{equation}
    \mathcal{R}= \xi^{\prime \, 2}\frac{m_1^2 x^2}{Q^2 \chi^2}, \quad \hat{t}_{1}(y)=\frac{\hat{s}_1}{\hat{s}_1+m_2^2} \Delta^{\prime} (y-y_0), \quad  y_0=\frac{1}{2} \left( 1+ \frac{\Sigma_{++}+ \hat{s}_1}{ \Delta^{\prime}}\right)\, .
\end{equation}
The variable $\hat{s}_1$ is expressed as a function of $\xi^\prime$ in the following way: 
\begin{equation}
\hat{s}_1(\xi') = \frac{1 - \xi'}{2\xi'} \left[ (\Delta - \Sigma_{+-}) \xi' + \Delta + \Sigma_{+-} \right]\, .
\end{equation}
With this, we can proceed to calculate the NLO contribution to the inclusive structure functions at the hadron level. 
They are given by
\begin{equation}
   g_{i}^{\rm QS, \, NLO}(x, Q^2, m_1, m_2)=\frac{\alpha_s}{2 \pi } \int_\chi^1 \frac{d \xi^{\prime}}{\xi^{\prime}}\left[\Delta f\left(\frac{\chi}{\xi^{\prime}}, \mu^2\right) \Delta \hat{\mathcal{H}}_i^{\rm QS, \, NLO}\left(\xi^{\prime},Q^2,m_1, m_2\right)\right],
   \label{NLO QS contribution}
\end{equation}
where $\Delta f(x,\mu^2)$ is the polarized parton density of the incoming heavy quark at the factorization scale $\mu$. For completeness, we report the structure functions at leading order,
\begin{equation}
  \Delta \hat{\mathcal{H}}_{i}^{\mathrm{QS,\,LO}} = \mathcal{N}_{i}^{\rm QS, \, LO}\delta(1-\xi^\prime), \qquad
g_{i}^{\mathrm{QS,\,LO}}(x,Q^2,m_1,m_2) =  \mathcal{N}_{i}^{\rm QS, \, LO}  \Delta f(\chi), \nonumber\\ 
\end{equation}
with 
\begin{align}
    \mathcal{N}_{\{1,5,7\}}^{\text{QS, LO}} &= \{N_1, \, \Gamma N_5, \, \Gamma^2 N_7\} \,, \nonumber\\
    \mathcal{N}_4^{\text{QS, LO}} &= \chi(1+\bar{\mathcal{R}}) \Gamma^2 N_4 \,, \quad \mathcal{N}_6^{\text{QS, LO}} = \Gamma \left(N_6 - (\Gamma-1) N_7 \right)\,,
\end{align}
where 
$\overline{\mathcal{R}}= m_1^2 x^2/(Q^2 \chi^2)$ and
we have introduced the factor $\Gamma = (1+\bar{\mathcal{R}})/(1-\bar{\mathcal{R}})$.

\section{Boson-gluon fusion}

To compute the contribution of the boson-gluon fusion process
\begin{equation}
B^*(q) + g(p) \to Q_1(k_1) + Q_2(k_2) ,
\end{equation}
the initial step involves the amplitude factorization as follows: 
\begin{equation}
\mathcal{M}^{\rm GF} = \varepsilon^{\nu} (q)\varepsilon^{\mu}(p) \mathcal{M}_{\mu \nu}^{\rm GF}.
\end{equation}
Here, $\varepsilon^{\nu}(q)$ and $\varepsilon^{\mu}(p)$ denote the polarization vectors of the incoming particles. Specifically, $q$ represents the momentum of the gauge boson $B^*$, while the momentum $p$ corresponds to the incoming gluon $g$. Then, the partonic tensor with helicity dependency is generalized in the following way:
\begin{equation}
\tilde{W}_{\mu \nu}^{\rm GF}(h) = \mathcal{E}^{\alpha \beta} (p,h) \mathcal{M}_{\mu \alpha}^{\rm GF} (\mathcal{M}_{\nu \beta}^{\rm GF} )^*,
\end{equation}
where we introduced the polarization tensor \( \mathcal{E}^{\mu \nu}(p,h) \) defined as~\cite{bojak2000nloqcdcorrectionspolarized}
\begin{equation}
\mathcal{E}^{\mu \nu}(p, h) = \varepsilon^\mu(p, h) \varepsilon^{\nu *}(p, h) =\frac{1}{2} \left( -g^{\mu \nu} + i h \epsilon^{\mu \nu \rho \sigma} \frac{p_\rho q_\sigma}{p \cdot q} \right),
\end{equation}
with \( h  \in \{-1, 1\} \) being the helicity of the incoming gluon. The symmetric component of the polarization tensor accounts for the unpolarized contribution, while the antisymmetric component corresponds to the polarized contribution. It is important to note that the helicity-dependent term is the only one allowed by Lorentz invariance and gauge freedom~\cite{bojak2000nloqcdcorrectionspolarized}.

As done for the quark scattering case, we define the tensor 
\begin{equation}
\Delta \hat{W}_{\mu \nu}^{\rm GF}= \frac{1}{2}\left( \tilde{W}_{\mu \nu}^{\rm GF}(-1) -\tilde{W}_{\mu \nu}^{\rm GF} (1)\right),
\end{equation}
from which we obtain the semi-inclusive partonic structure functions 
\begin{equation}
\hat{g}_i^{\rm GF} \equiv \left( \frac{g_s^2}{2 \pi} \right)^{-1} \hat{P}_{i, \rm GF}^{\mu \nu} \overline{ \Delta \hat{W}_{\mu \nu}^{\rm GF}}, \quad i \in \{1, 4, 5, 6, 7\}.
\end{equation}
Here, the averaged partonic tensor is given by
\begin{equation}
\overline{ \Delta \hat{W}_{\mu \nu}^{\rm GF}} = \frac{1}{2} \times \frac{1}{8} \sum_{\rm color} \Delta \hat{W}_{\mu \nu}^{\rm GF},
\end{equation}
and the factor of 8 comes from averaging over the color states of the incoming gluon.
The inclusion of the additional $1/2$ factor in the GF structure functions is dictated by the requirement of consistency with the leading order result, reflecting the standard normalization~\cite{Aschenauer_2013,bierenbaum$Oa_s^2$PolarizedHeavy2023}. This is because the GF process in the collinear limit must factorize into the polarized splitting function and the LO structure function, so the same normalization must be preserved.

\begin{figure}[H]
    \centering
    \includegraphics[scale=0.8]{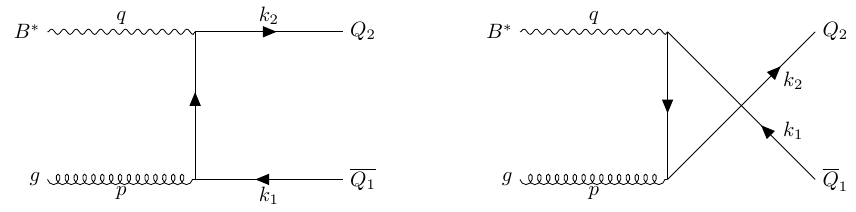}
    \caption{Representative real-emission diagrams for boson-gluon fusion, $B^*(q) + g(p) \to Q_1(k_1) + Q_2(k_2)$, at $\mathcal{O}(\alpha_s)$. These diagrams contribute to the polarized partonic structure functions $\hat g_i^{\rm GF}$ and provide the gluon-initiated component of the NLO coefficient functions entering the ACOT scheme.}
    \label{fig: Boson Gluon fusion}
\end{figure}

To compute the relevant contributions, we must consider the Feynman diagrams associated with the real contribution at next-to-leading order for the boson-gluon fusion process. These are illustrated in~\cref{fig: Boson Gluon fusion}.
For convenience, we introduce the variables
\begin{equation}
\zeta = \frac{p \cdot k_2}{p \cdot q}, \quad \hat{s} = (q + p)^2 = Q^2 \left( \frac{1}{\hat{x}} - 1 \right),
\end{equation}
where \( \hat{x} \) is the Bjorken-like variable, and we calculate the process using  the convention \( \hat{u} = (k_2 - p)^2 \).
The semi-inclusive partonic structure functions are decomposed as
\begin{equation}
\hat{g}_i^{\rm GF} =   4 \pi \mathcal{C}_i \left( \frac{A_i}{(1 - \zeta)^2} + \frac{B_i}{\zeta^2} + \frac{C_i}{1 - \zeta} + \frac{D_i}{\zeta} + E_i \right),
\label{Boson Gluon Fusion semi-inclusive}
\end{equation}
where the coefficients \( A_i \), \( B_i \), \( C_i \), \( D_i \), and \( E_i \) are given by
\begin{align}
A_1 &= \frac{1}{2}   \hat{x} \frac{m_1^2}{Q^2}, & C_1 &= \frac{1}{2} \left(\Delta P_{qg}^{(0)} \left( \frac{\hat{x}}{\lambda}\right)  -2   A_1\right), & E_1 &= -\Delta P_{qg}^{(0)}(\hat{x}), \nonumber\\
A_4 &= 2 \hat{x} (A_5 + A_6), & C_4 &= 2 \hat{x} (C_5 + C_6), & E_4 &= 0,\nonumber \\
A_5 &= -2 A_1, & C_5 &= -2 C_1, & E_5 &= 0, \nonumber\\
A_6 &= \frac{m_2^2 - m_1^2}{Q^2} A_5, & C_6 &= \frac{m_2^2 - m_1^2}{Q^2} (C_5 + 2 A_5), & E_6 &= 0, \nonumber\\
A_7 &= -2 A_6, & C_7 &= -2 C_6, & E_7 &= 0,
\end{align}
with $\lambda= Q^2/(m_2^2+Q^2)$ and the coefficients \( B_i \) and \( D_i \) defined as
\begin{equation}
B_{\substack{i = 1}}^{\, 4, 5, 6, 7}= \mp A_i \left( m_1 \leftrightarrow m_2 \right), \quad D_{\substack{i = 1}}^{\, 4, 5, 6, 7} = \mp C_i \left( m_1 \leftrightarrow m_2 \right).
\label{eq: bgf symmetry}
\end{equation}
The factors $\mathcal{C}_i$ are expressed in terms of the axial and vector coupling combinations $q_+ =  V'V+AA'$ and $R_+ = (AV' + A'V)/2$:
\begin{equation}
\mathcal{C}_1 = q_+, \quad \mathcal{C}_i = R_+ \quad \text{for } i = 4, 5, 6, 7\ ,
\end{equation}
where $q_+$ and $R_+$ are defined in~\cref{s1 expression}. The quark-gluon polarized splitting function \( \Delta P_{qg}^{(0)}(\hat{x}) \) is defined as~\cite{altarelliAsymptoticFreedomParton1977a}
\begin{equation}
\Delta P_{qg}^{(0)}(\hat{x}) =  \frac{1}{2} \left(2 \hat{x} - 1\right).
\end{equation}
Given the properties reported in~\cref{eq: bgf symmetry}, and considering the parity-violating nature of these structure functions, boson-gluon fusion does not contribute to the neutral-current structure functions. Instead, it contributes only to the charged-current structure functions when the full massive contributions are taken into account. This follows from charge-conjugation symmetry.

Finally, we note that the Slavnov-Taylor identities~\cite{Slavnov:1972fg} and the structure of the projectors in~\cref{Projectors appendix} imply the following relations at every order in perturbation theory:
\begin{equation}
\hat{g}_{4}^{\rm GF}= 2 \hat{x}( \hat{g}_5^{\rm GF}+\hat{g}_{6}^{\rm GF}) , \quad \hat{g}_{7}^{\rm GF}=-2 \hat{g}_{6}^{\rm GF}. 
\label{Boson Gluon relations}
\end{equation}

It is important to note, however, that in the quark-scattering channel, the relation linking $g_4$ to the structure functions $g_5$ and $g_6$ is violated, even in the massless limit. This violation is physically significant and corresponds to the breakdown of the Dicus relation~\cite{PhysRevD.5.1367} at next-to-leading order. This effect is the polarized analog of the violation of the Callan-Gross relation observed in unpolarized structure functions.

Considering the NC case, our results for the structure function $g_{\{1, \, 4, \, 5\}}$ are in full agreement with previously published expressions in the literature~\cite{Hekhorn:2019nlf}. 
The Wilson coefficients are then given by
\begin{equation}
\Delta\mathcal{H}_4^{\rm GF} = \frac{x}{\hat{x}} \mathfrak{g}_4^{\rm GF}, \quad \Delta \mathcal{H}_i^{\rm GF} = \mathfrak{g}_i^{\rm GF}, \quad \text{for} \,  i= 1,5,6,7\ ,
\end{equation}
where
\begin{equation}
    \mathfrak{g}_i^{\rm GF}(\hat{x},Q^2) \equiv  \frac{1}{8 \pi}\int_{\zeta_-}^{\zeta_+} \hat{g}_i^{\rm GF}(\zeta,\hat{x},Q^2) \, d \zeta,
\end{equation}
and the variables \( \zeta_{\pm} \) are defined as
\begin{equation}
    \begin{aligned}
        &\zeta_{\pm} = \dfrac{1}{2}\left(\zeta_1 \pm \zeta_2\right), \\
        &\zeta_1 = 1 + \dfrac{m_2^2 - m_1^2}{\hat{s}}, \quad \zeta_2 = \dfrac{1}{\hat{s}}\Delta(\hat{s},m_2^2,m_1^2).
    \end{aligned}
\end{equation}
Finally, the next-to-leading order contribution from the boson-gluon fusion process is given by
\begin{equation}
g_i^{\rm GF, \, NLO} = \frac{\alpha_s}{2 \pi} \Delta \mathcal{H}_i^{\rm GF, \, NLO} \otimes \Delta g,
\end{equation}
where $\Delta g$ stands for the polarized gluon PDF. Explicitly, it takes the form
\begin{equation}
g_i^{\rm GF, \, NLO}(x,Q^2,m_1,m_2) = \frac{\alpha_s}{2 \pi} \int_{\chi'}^1 \frac{d \xi'}{\xi'} \left[   \Delta g\left( \frac{\chi'}{\xi'}, \mu^2 \right) \Delta\hat{\mathcal{H}}_i^{\rm GF} \left( \xi', \frac{m_1}{Q}, \frac{m_2}{Q} \right) \right],
\label{GF convolution}
\end{equation}
where the variables are related by
\begin{equation}
  \xi^\prime= \frac{\chi^\prime}{x} \hat{x} , \quad \chi' = x \frac{(m_1 + m_2)^2 + Q^2}{Q^2}.
\end{equation}

\section{Structure functions in the ACOT scheme}

The theoretical treatment of heavy quarks in deep-inelastic scattering requires a consistent framework that captures mass effects near the production threshold while resumming large quasi-collinear logarithms at asymptotically high scales \cite{Aivazis:1993kh, Aivazis:1993pi, Kretzer:1998nt}. To achieve this, we employ the ACOT scheme, which provides a systematic method for incorporating heavy-quark mass effects within a variable flavor number scheme. The conceptual core of the ACOT scheme lies in its treatment of heavy quarks within the factorization framework: the heavy-quark mass acts as a physical regulator for collinear behavior near the threshold, while the factorization scale $\mu$ controls the transition to a regime where the heavy quark is treated as an active parton with its own distribution function. To maintain perturbative consistency and avoid double-counting the resummed logarithms already contained in the heavy-quark PDFs, the ACOT scheme introduces explicit subtraction terms. These terms remove the contributions reproduced by the zero-mass limit from the massive next-to-leading order coefficient functions. Furthermore, we adopt the CWZ decoupling prescription \cite{PhysRevD.18.242}, which ensures that the subtraction is only active when the factorization scale $Q$ exceeds the heavy-quark mass.

In this work, we present the explicit subtraction terms adopting the conventional scale choice $\mu_{R} = \mu_{F} = Q$. This choice serves to simplify the analytic structure of the coefficients by eliminating explicit renormalization scale logarithms in the definition. While modern phenomenological applications sometimes employ more general dynamical scales to improve perturbative stability near thresholds, the standard choice ($\mu_{R} = \mu_{F} = Q$) remains the foundation for the original ACOT formulation.

As is well known, the structure functions within the ACOT scheme are defined by the following expression, which explicitly accounts for the necessary subtraction terms:
\begin{equation}
    g_{i}^{\rm ACOT}= g_{i}^{\rm QS, \, LO}+ g_{i}^{\rm QS, \, NLO}+ g_{i}^{\rm GF,\, NLO}- g_{i}^{\rm QS, \, SUB}- g_{i}^{\rm GF, \, SUB}, \quad i=1, 4,5,6,7.
\end{equation}
For further details, we refer the reader to Refs.~\cite{Aivazis:1993kh,Aivazis:1993pi,Kretzer:1998ju,spezzanoHeavyquarkContributionsDIS2025}.

The practical implementation of the ACOT scheme necessitates the explicit computation of subtraction terms. As prescribed by the formalism, these terms fulfill a dual purpose: first, they must cancel the double-counting contributions near the heavy-quark threshold; second, they must regularize any emergent collinear logarithms. Furthermore, in the high-energy limit, the results are constrained to reproduce the standard $\overline{\mathrm{MS}}$ coefficients.
In particular, the gluon fusion subtraction term is given by
\begin{align}
    g_i^{\rm GF, \, SUB}\left(x, Q^2, m_1, m_{2}\right) &= \mathcal{N}_i^{\rm QS, \, LO} \frac{\alpha_s(Q^2)}{2\pi}
    \int_{\chi}^1 \frac{\mathrm{d}\xi'}{\xi'} \,
    \Delta g\!\left(\frac{\chi}{\xi'},Q^2\right) \,
    \Delta \mathcal{H}_i^{\rm GF, \,SUB}\left(\xi', \frac{m_1}{Q}, \frac{m_2}{Q}\right),
\end{align}
 where the indices are $i \in \{1, 4, 5, 6, 7\}$ and  the kernels are given by:
\begin{align}
    \Delta \mathcal{H}_1^{\rm GF, \,SUB}\left(\xi', \frac{m_1}{Q}, \frac{m_2}{Q}\right) &= \frac{1}{2}\Delta P_{q g}^{(0)}(\xi^\prime)\left[ \Theta \left(1-\frac{m_1^2}{Q^2}\right) \ln\frac{Q^2}{m_1^2}
    + \Theta \left(1-\frac{m_2^2}{Q^2}\right) \ln\frac{Q^2}{m_2^2} \right], \nonumber\\[8pt]
    \Delta \mathcal{H}_{4,5,6,7}^{\rm GF, \, SUB}
    \left(
       \xi^\prime,
       \frac{m_1}{Q}, \frac{m_2}{Q}
    \right)
    &=\frac{1}{2}
    \Delta P_{q g}^{(0)}(\xi^\prime) \left[ \Theta \left(1-\frac{m_2^2}{Q^2}\right) \ln\frac{Q^2}{m_2^2}
    -\Theta \left(1-\frac{m_1^2}{Q^2}\right) \ln\frac{Q^2}{m_1^2} \right].
\end{align}
Here, $\Theta(x)$ denotes the Heaviside step function. The minus sign in the subtraction terms for the parity-violating structure functions is a direct consequence of the axial-vector coupling and the charge/flavor structure of the weak interaction. Conversely, for the quark-scattering channel, we derive the relations
\begin{equation}
    g_i^{\rm QS,\,SUB}\left(x, Q^2, m_1, m_{2}\right) = \mathcal{N}_i^{\rm QS,\,LO} \frac{\alpha_s(Q^2)}{2\pi}
    \int_{\chi}^1 \frac{\mathrm{d}\xi'}{\xi'} \,
    \Delta f\!\left(\frac{\chi}{\xi'},Q^2\right) \,
    \Delta \mathcal{H}_i^{\rm QS,\,SUB}\left(\xi', \frac{m_1}{Q}\right),
\end{equation}
where the subtraction kernel is defined as
\begin{equation}
    \Delta \mathcal{H}_{i}^{\rm QS,\,SUB}\!\left(\xi^\prime, \frac{m_1}{Q}\right)
    = c_i  C_F\left(\left[\frac{1+\xi^{\prime 2}}{1-\xi^\prime}
    \left(
    \ln\frac{Q^2}{m_1^2}
    -2\ln(1-\xi^\prime)
    -1
    \right)\right]_+ -2(1-\xi^\prime)\right)
    \label{eq:sub}
\end{equation}
with coefficients $c_{1,4,5}=1$ and $c_{6,7}=0$.  

Finally, taking the massless limit, for both channels, we find the asymptotic expressions
\begin{align}
    \lim_{m_2 \to 0} \lim_{m_1 \to 0} \frac{1}{\mathcal{N}_i}\left[\Delta \mathcal{H}_{i}^{\rm GF, \, NLO} \left(\xi^\prime,\frac{m_1}{Q}, \frac{m_2}{Q} \right)-\Delta \mathcal{H}_{i}^{\rm GF, \, SUB} \left(\xi^\prime,\frac{m_1}{Q}, \frac{m_2}{Q}  \right) \right] &= \Delta C_{g,i}(\xi^\prime),\nonumber \\
    \lim_{m_2 \to 0} \lim_{m_1 \to 0} \frac{1}{\mathcal{N}_i}\left[\Delta\mathcal{H}_{i}^{\rm QS, \, NLO} \left(\xi^\prime, \frac{m_1}{Q}, \frac{m_2}{Q} \right)-\Delta \mathcal{H}_{i}^{\rm QS, \, SUB} \left(\xi^\prime,\frac{m_1}{Q}, \frac{m_2}{Q}  \right) \right] &= \Delta C_{q,i}(\xi^\prime)
\end{align}
with 
\begin{equation}
    \mathcal{N}_1= \frac{1}{2} , \quad \mathcal{N}_4=2x, \quad \mathcal{N}_{5,6,7}=1.
\end{equation}
The normalization factors $\mathcal{N}_i$ are the necessary corrections in order to maintain a unified convolution structure across all indices (cf.~\cref{NLO QS contribution,GF convolution}).
With this choice, our expressions exactly reproduce the standard results in the literature (e.g.,~\cite{Aschenauer_2013}); the difference is purely notational. In particular, we factor out index-specific kinematic conventions, such as $1/2$ for $g_1$ and $2x$ for $g_4$, that are traditionally embedded within the definition of each convolution. This ensures that the hard kernels $\Delta \mathcal{H}_i$ map directly onto the standard coefficient functions $\Delta C_i$ in the massless limit, providing an important validation of both the calculation and the subtraction framework.

The coefficient functions have already been computed in the literature~\cite{Vogelsang_1996, Aschenauer_2013}. For completeness, we report here their representation in $\xi'$-space:
\begin{align}
   \Delta C_{q,1}(\xi^\prime) &= C_F \left(-\frac{3}{2} \delta(1-\xi^\prime) + \left[ \frac{1 + \xi^{\prime 2}}{1 - \xi^\prime} \ln\left(\frac{1 - \xi^\prime}{\xi^\prime}\right) - \frac{3}{2(1 - \xi^\prime)} + \xi^\prime + 2 \right]_+ \right),\nonumber\\
    \Delta C_{q,4}(\xi^\prime) &= C_F \left[ \frac{1 + \xi^{\prime 2}}{1 - \xi^\prime} \ln\left(\frac{1 - \xi^\prime}{\xi^\prime}\right) - \frac{3}{2(1 - \xi^\prime)} + 2\xi^\prime + 3 \right]_+ ,\nonumber\\
   \Delta C_{q,5}(\xi^\prime) &= C_F\left( -\delta(1-\xi^\prime) + \left[ \frac{1 + \xi^{\prime 2}}{1 - \xi^\prime} \ln\left(\frac{1 - \xi^\prime}{\xi^\prime}\right) - \frac{3}{2(1 - \xi^\prime)} + 3 \right]_+\right),\nonumber\\
   \Delta C_{q,6,7}(\xi^\prime)&=0,\nonumber\\
    \Delta C_{g,1}(\xi^\prime)
&= \frac{1}{2} \left[ (2\xi^\prime - 1)\,\ln\!\left(\frac{1-\xi^\prime}{\xi^\prime}\right) + 3 - 4\xi^\prime \right],\nonumber\\
\Delta C_{g,4,5,6,7}(\xi^\prime) &=0 .
\end{align}
Note that in Ref.~\cite{Aschenauer_2013} the coefficient functions are written in Mellin space and with an overall prefactor $\alpha_s/(2\pi)$ by convention. For completeness, we also recall the Mellin transform relation
\begin{equation}
    \Delta C_{q/g, i}(n) = \int_{0}^{1} \mathrm{d}\xi' \, {\xi'}^{\,n-1} \, \Delta C_{q/g, i}(\xi'),
\end{equation}
which is useful for switching between the $\xi'$-space and Mellin-$n$ representations.

A notable feature of the polarized framework, distinguishing it from the unpolarized case, is its sensitivity to the helicity-flip contributions of the heavy quark. In the massive calculation, these contributions do not decouple in the massless limit but instead leave a finite remainder. This effect is accounted for by the inclusion of the $-2(1-\xi^\prime)$ term in the subtraction kernel~\cref{eq:sub}. Physically, this term represents the manifestation of the well-known $\gamma_5$ artifact in the time-like quark scattering channel. As discussed in Ref.~\cite{Vogelsang_1996}, the application of dimensional regularization using prescriptions such as HVBM~\cite{THOOFT1972189} leads to an artificial helicity non-conservation at the quark-gluon vertex in $d \neq 4$ dimensions. While this manifests as a violation of the non-singlet axial current conservation in the space-like case, in the time-like case it appears as a finite discrepancy in the coefficient functions. The subtraction of $2(1-\xi^\prime)$ thus serves as a finite renormalization required to restore helicity conservation and ensure consistency with the standard $\overline{\mathrm{MS}}$ results.

\section{Numerical implementation}
\label{sec:numerical_results}

In this section, we present the numerical evaluation of the polarized structure functions for heavy-quark production. Our implementation is developed in \texttt{Mathematica}, using the \texttt{ManeParse} package~\cite{clarkManeParseMathematicaReader2017} to interface with the \texttt{LHAPDF} library~\cite{Buckley_2015}. For the results presented below, we focus on EIC phenomenology, assuming that the incoming lepton is an electron.

The electroweak sector is treated within the on-shell renormalization scheme. The primary input parameters---the fine-structure constant $\alpha$, the $W$-boson mass $M_W$, and the $Z$-boson mass $M_Z$---are fixed to their measured physical (pole) values from the Particle Data Group~\cite{ParticleDataGroup:2022pth}:
\begin{equation}
M_W = 80.3692~\text{GeV}, \qquad M_Z = 91.1876~\text{GeV}.
\end{equation}
In this scheme, the weak mixing angle is defined as a constant via the tree-level relation $\sin^2 \theta_W = 1 - M_W^2/M_Z^2$, neglecting any explicit scale dependence in $\sin^2 \theta_W$ or $\alpha$. This fixed-coupling approach is consistent with the treatment adopted in the \texttt{NNPDFpol1.1} polarized parton distributions~\cite{Nocera:2014gqa}, where electroweak running effects are neglected in the global fits. Given the typical EIC kinematic range, where the squared momentum transfer $Q^2$ remains well below $M_Z^2$, the impact of EW evolution is subdominant compared to QCD uncertainties and the running of the strong coupling $\alpha_s(Q^2)$.

The electromagnetic coupling is fixed at the low-energy value $\alpha = 1/137.035$~\cite{ParticleDataGroup:2022pth}. For the strong coupling, we employ running at NLO to maintain consistency with the perturbative order of the coefficient functions in this work, with the value at the $Z$-pole taken as $\alpha_s(M_Z) = 0.1180$~\cite{ParticleDataGroup:2022pth}. The CKM matrix elements used in the numerical analysis are taken from~\cite{ParticleDataGroup:2022pth}:
\begin{equation}
\left|V_{\mathrm{CKM}}\right| = 
\begin{pmatrix}
|V_{ud}| & |V_{us}| & |V_{ub}| \\
|V_{cd}| & |V_{cs}| & |V_{cb}| \\
|V_{td}| & |V_{ts}| & |V_{tb}|
\end{pmatrix}= 
\begin{pmatrix}
0.97435 & 0.22501 & 0.003732 \\
0.22487 & 0.97349 & 0.04183 \\
0.00858 & 0.04111 & 0.999118
\end{pmatrix}.
\end{equation}
Finally, the heavy-quark masses are fixed according to the prescriptions of the chosen PDF sets: $m_c = 1.41~\text{GeV}$, $m_b = 4.75~\text{GeV}$, and $m_t = 175~\text{GeV}$.

\subsection{Numerical results and phenomenological discussion}
\label{sec:implementation}

Having established the theoretical framework for the polarized ACOT scheme, we now turn to its numerical implementation and a discussion of the resulting phenomenological implications. The inclusion of heavy-quark mass effects is essential for the reliable description of polarized DIS observables, particularly in the vicinity of flavor thresholds where the zero-mass approximation is expected to be inadequate. Such a consistent treatment is a prerequisite for the high-precision requirements of future experimental programs, most notably the Electron-Ion Collider (EIC).

The definition of the NC and CC structure functions was already outlined in \cref{sec:kinematics}, i.e.
\begin{align}
 &g_{\{1,4,5\}}^{\text{CC}} = g_{\{1,4,5\}}^{W^\pm}, \nonumber\\
 &g_1^{\text{NC}} = g_1^\gamma - g_V^e \eta_{\gamma Z} g_1^{\gamma Z} + \left( g_V^{e \, 2} + g_A^{e \, 2} \right) \eta_Z g_1^Z, \nonumber\\
 &g_{\beta}^{\text{NC}} = g_A^e \eta_{\gamma Z} g_\beta^{\gamma Z} - 2 g_V^e g_A^e \eta_Z g_\beta^Z,\quad \beta= 4, \, 5,
\end{align}
where the couplings $g_V^e$, $g_A^e$ and the propagator factors $\eta_j$ were given in \cref{eq:couplings,eq:eta factors}.
The implementation of the massive structure functions in the ACOT scheme is carried out channel by channel to correctly account for the flavor and mass dependencies in NC and CC interactions. Following the convention in Ref.~\cite{spezzanoHeavyquarkContributionsDIS2025}, we define
\begin{equation}
  \Delta\mathcal{H}_{i,\,\alpha,\,\beta}^{\mathrm{QS/GF},\,j}(\xi') \equiv \Delta \mathcal{H}_{i}^{\mathrm{QS/GF}} \!\left(\xi', \frac{m_\alpha}{Q}, \frac{m_\beta}{Q}\right),
  \label{eqn:index_convention_coefficients}
\end{equation}
where $\alpha,\beta \in \{u,d,s,c,b,t\}$ and $j \in \{\gamma, \gamma Z, Z, W\}$. The vector and axial couplings $V, V', A,$ and $A'$ for each channel are summarized in \cref{tab:couplings_full}.
\begin{table}[tbp]
\caption{Electroweak couplings for the various interaction channels and quark families.}
\centering
\renewcommand{\arraystretch}{1.2}
\begin{tabular}{|l|l|cccc|}
\hline
Channel ($j$) & Flavor Type & $V$ & $V'$ & $A$ & $A'$ \\
\hline
Photon ($\gamma$) & up-type & $2/3$ & $2/3$ & $0$ & $0$ \\
 & down-type & $-1/3$ & $-1/3$ & $0$ & $0$ \\
\hline
Interference ($\gamma Z$) & up-type & $2/3$ & $\frac{1}{2} - \frac{4}{3}\sin^2\theta_W$ & $0$ & $1/2$ \\
 & down-type & $-1/3$ & $-\frac{1}{2} + \frac{2}{3}\sin^2\theta_W$ & $0$ & $-1/2$ \\
\hline
$Z$-Boson ($Z$) & up-type & $\frac{1}{2} - \frac{4}{3}\sin^2\theta_W$ & $\frac{1}{2} - \frac{4}{3}\sin^2\theta_W$ & $1/2$ & $1/2$ \\
 & down-type & $-\frac{1}{2} + \frac{2}{3}\sin^2\theta_W$ & $-\frac{1}{2} + \frac{2}{3}\sin^2\theta_W$ & $-1/2$ & $-1/2$ \\
\hline
$W$-Boson ($W$) & all-type & $1$ & $1$ & $1$ & $1$ \\
\hline
\end{tabular}
\label{tab:couplings_full}
\end{table}
The ACOT scheme results from the combination of massive contributions at LO, NLO, and the corresponding subtraction terms, i.e.
\begin{align}
    g_{i,\, \alpha,\, \beta}^{\mathrm{QS, \, ACOT},\,j} &= g_{i,\,\alpha,\,\beta}^{\mathrm{QS,\, LO},\, j} + g_{i,\,\alpha,\,\beta}^{\mathrm{QS,\, NLO},\, j} - g_{i,\, \alpha,\, \beta}^{\mathrm{QS,SUB},j}, \nonumber\\
    g_{i,\,\alpha,\,\beta}^{\mathrm{GF, \,ACOT},\, j} &= g_{i,\,\alpha,\,\beta}^{\mathrm{GF,\,NLO},\,j} - g_{i,\,\alpha,\,\beta}^{\mathrm{GF,\,SUB},\,j}.
\end{align}
Summing over the up-type ($U$) and down-type ($D$) sectors for each NC channel $\bar{j} \in \{\gamma, \gamma Z, Z\}$, we obtain
\begin{equation}
    g_i^{\bar{j},\, \mathrm{ACOT}} = \sum_{q \in \{U,D\}} \Bigl[ g_{i,q,q}^{\mathrm{QS,ACOT},\bar{j}} + g_{i,\bar q,\bar q}^{\mathrm{QS,ACOT},\bar{j}} + 2\,g_{i,q,q}^{\mathrm{GF,ACOT},\bar{j}} \Bigr].
    \label{eqn:NC_ACOT_full}
\end{equation}
The gluon-fusion (GF) term appears only for $g_1$; for the parity-violating structure functions, it vanishes identically. For the CC sector, the expressions are
\begin{subequations}
\begin{align}
    g_i^{W^+,\,\mathrm{ACOT}} &= \sum_{U,D}|V_{UD}|^2 \Bigl[ g_{i,\bar D,\bar U}^{\mathrm{QS,ACOT},W} + g_{i,U,D}^{\mathrm{QS,ACOT},W} + g_{i,\bar D,\bar U}^{\mathrm{GF,ACOT},W} + g_{i,U,D}^{\mathrm{GF,ACOT},W} \Bigr], \\
    g_i^{W^-,\,\mathrm{ACOT}} &= \sum_{U,D}|V_{UD}|^2 \Bigl[ g_{i,D,U}^{\mathrm{QS,ACOT},W} + g_{i,\bar U,\bar D}^{\mathrm{QS,ACOT},W} + g_{i,D,U}^{\mathrm{GF,ACOT},W} + g_{i,\bar U,\bar D}^{\mathrm{GF,ACOT},W} \Bigr].
\end{align}
\end{subequations}
For the numerical treatment of the plus-distribution, we remind the reader that the full details of the implementation and the regularization are provided in the appendices C and D of Ref.~\cite{spezzanoHeavyquarkContributionsDIS2025}. Our results focus on the kinematic region $x \in [0.01, 0.9]$ and $Q^2 \in [m_c^2, 10^4]\, \mathrm{GeV}^2$. 

\begin{figure}[tbp]
    \centering
    \includegraphics[width=\textwidth]{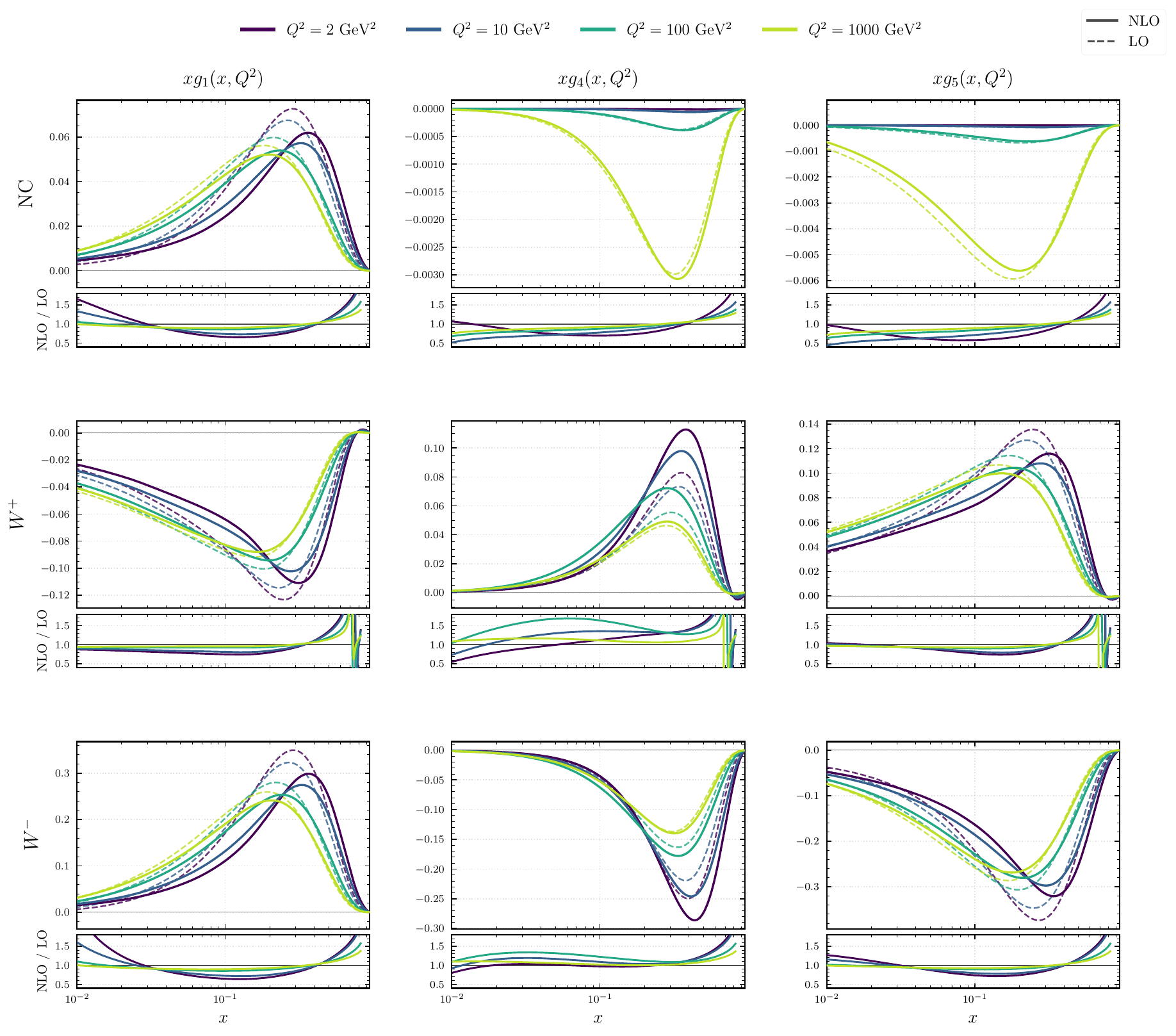}
    \caption{Polarized structure functions $x g_1$, $x g_4$, and $x g_5$ in the ACOT scheme for NC and CC DIS. Results are shown at LO (dashed) and NLO (solid) for representative scales $Q^2 \in \{2,\,10,\,100,\,1000\}~\mathrm{GeV}^2$. The lower panels display the ratio $\text{NLO}/\text{LO}$, highlighting the impact of QCD corrections.}
    \label{fig:structure_functions}
\end{figure}

\begin{figure}[tbp]
    \centering
    \includegraphics[width=0.87\textwidth]{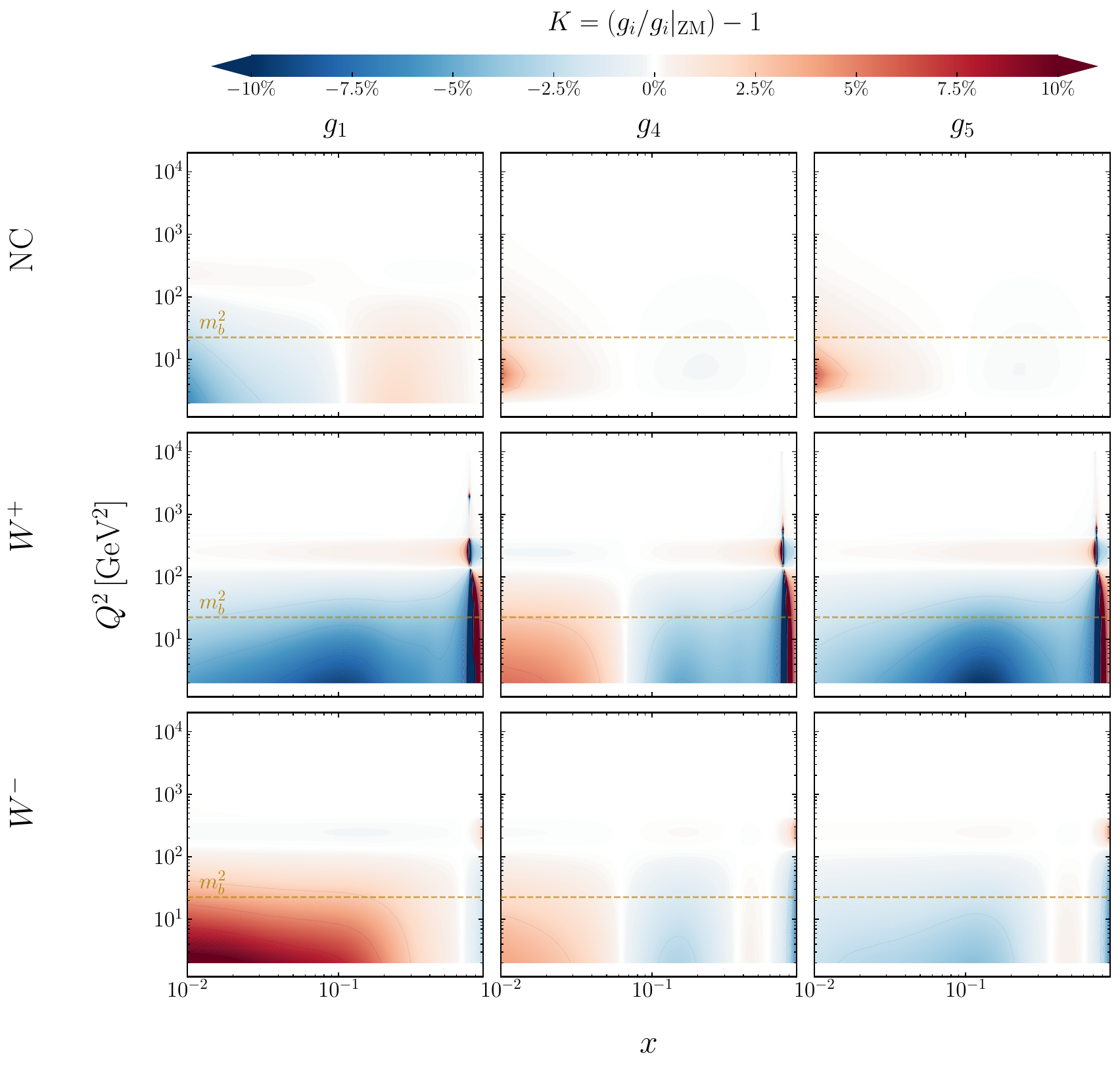}
    \caption{Heatmaps of the relative deviation $K = g_i / g_i^{\rm ZM} - 1$. The color scale indicates the percentage deviation in the $(x,Q^2)$ plane. The dashed line marks the bottom-quark threshold $Q^2 \simeq m_b^2$.}
    \label{fig:heatmap}
\end{figure}

Numerical results for the structure functions are illustrated in \cref{fig:structure_functions,fig:heatmap}.
As seen in \cref{fig:structure_functions}, the LO structure functions approach zero in the high-$x$ limit, yet the $K$-factors show a significant enhancement as $x \to 1$. This behavior is a signature of Sudakov logarithm effects~\cite{Catani:1989ne} arising from soft-gluon emissions near the kinematic boundary. 
Furthermore, the heatmaps in \cref{fig:heatmap} quantify the relative deviation between the ACOT results and the zero-mass (ZM) baseline, $K = (g_i / g_i|_{\text{ZM}}) - 1$. 
We observe variations of up to 10\% for scales $Q^2$ near or below the bottom-quark threshold, particularly at moderate $x$. This highlights the failure of the ZM approximation to account for the restricted phase space of massive final states. In the high-$Q^2$ region, the heavy-quark mass contributions decouple, and the ACOT results converge to the ZM limit as expected. 
Notably, in the $W^+$ channel at high $x$, the heatmap displays a localized enhancement. This is a numerical artefact caused by the ZM structure function crossing zero in this region, resulting in an unphysical divergence in the ratio $K$ rather than a purely dynamical effect. 

In summary, these results provide a critical baseline for the extraction of polarized parton distributions at future facilities, demonstrating that a massive treatment is essential near flavor thresholds and in the large-$x$ regime.

\clearpage
\section{Conclusions and outlook}
\label{sec:conclusions}

In this work, we have presented a comprehensive calculation of heavy-quark contributions to the polarized deep-inelastic scattering structure functions $g_1, g_4, g_5, g_6$, and $g_7$ at next-to-leading order in perturbative QCD. Working within the ACOT general-mass variable-flavor-number scheme, we derived closed analytic expressions for the massive quark-scattering and boson--gluon-fusion coefficient functions. Crucially, our formalism retains the full dependence on both initial- and final-state quark masses, providing a consistent framework for describing heavy-flavor production across all energy scales.

From a phenomenological perspective, we implemented these massive coefficient functions and the corresponding subtraction terms in a dedicated \texttt{Mathematica} code, interfaced with modern polarized parton distribution functions via the \texttt{ManeParse} framework. Focusing on the kinematics of the future Electron Ion Collider, we performed a detailed comparison between LO and NLO predictions, as well as between the full ACOT results and the standard zero-mass approximation for both neutral-current and charged-current channels.

Our numerical analysis revealed that NLO QCD corrections induce sizable modifications to the LO structure functions, characterized by large $K$-factors typically driven by Sudakov-like logarithms. Near heavy-quark production thresholds, the massive treatment leads to a substantial suppression of the structure functions relative to the ZM limit, reflecting the physical constraints of the reduced phase space. Specifically, the full massive NLO corrections can reach variations of $\mathcal{O}(5-10\%)$ in the kinematic regions close to or below the bottom-quark threshold, underscoring the necessity of a rigorous massive treatment for high-precision phenomenology.

While we adopted the conventional scale choice $\mu_R = \mu_F = Q$ throughout this study, our methodology is inherently flexible and can be readily extended to independent scale variations. Such investigations will be essential for future global analyses to quantify theoretical uncertainties. At high scales $Q^2 \gg m_b^2$, the ACOT results smoothly recover the zero-mass predictions, confirming the correct asymptotic behavior of the scheme.

The results provided here offer a theoretically consistent and numerically robust description of heavy-flavor effects in polarized DIS, suitable for the precision requirements of future facilities such as the EIC and EicC. These results can be directly integrated into global PDF fits to better constrain the polarized gluon distribution and heavy-quark helicity densities. Several directions for future work include the incorporation of NNLO corrections, threshold and small-$x$ resummations, and the investigation of power-suppressed higher-twist effects relevant for $g_2$ and $g_3$. Furthermore, interfacing our results with public PDF fitting codes and Monte Carlo frameworks will be a vital step toward a quantitatively reliable spin-physics program at next-generation lepton--hadron colliders.

The structure functions $g_{6,7}$ are suppressed by the lepton mass squared in the cross sections and we therefore didn't show numerical results here. They may, however, become relevant in polarized DIS with third generation leptons and a phenomenological study could be carried out in this case in the future.

\appendix
\clearpage
\section{Projectors}
\label{Projectors appendix}

Considering the expression for the hadronic tensor
\begin{align}
W_{\mu \mu^{\prime}}(S) = & \,{-g_{\mu \mu^{\prime}} F_1 + \frac{P_{ \mu} P_{ \mu^{\prime}}}{P \cdot q} F_2} - i \varepsilon_{\mu \mu^{\prime} \alpha \beta} \frac{q^\alpha P^\beta}{2 P \cdot q} F_3 + \frac{q_\mu q_{\mu^{\prime}}}{Q^2} F_4 + \frac{q_\mu P_{ \mu^{\prime}} + q_{\mu^{\prime}} P_{ \mu}}{2 P \cdot q} F_5 \nonumber\\
& + i \varepsilon_{\mu \mu^\prime \alpha \beta} \frac{q^\alpha}{P \cdot q}\left[S^\beta g_1 + \left(S^\beta - \frac{S \cdot q}{P \cdot q} P^\beta \right) g_2\right] + \frac{1}{P \cdot q}\left[\frac{1}{2} \left(P_{ \mu}^\perp S_{\mu^\prime}^\perp + S_\mu^\perp P_{ \mu^\prime}^\perp\right) - \frac{S \cdot q}{P \cdot q} P_{ \mu}^\perp P_{ \mu^\prime}^\perp \right] g_3 \nonumber \\
& + \frac{S \cdot q}{P \cdot q}\left[\frac{P_{ \mu}^\perp P_{ \mu^{\prime}}^\perp}{P \cdot q} g_4 + \left(-g_{\mu \mu^{\prime}} + \frac{q_\mu q_{\mu^{\prime}}}{q^2}\right) g_5\right] + \frac{S \cdot q}{P \cdot q}\left[\frac{q_\mu q_{\mu^{\prime}}}{q^2} g_6 + \frac{q_\mu P_{ \mu^{\prime}} + q_{\mu^{\prime}} P_{ \mu}}{2 P \cdot q} g_7\right],
\end{align}
where 
\begin{equation}
    P_{ \mu}^\perp = P_{\mu} - \frac{P \cdot q}{q^2} q_\mu, \quad S_\mu^\perp = S_\mu - \frac{S \cdot q}{q^2} q_\mu,
     \label{partonic tensor}
\end{equation}
the following projectors can be derived:
\begin{align}
  P_{1}^{\mu \nu}(P,q) &= -\frac{i}{\Delta} \varepsilon^{\mu \nu \alpha \beta} P_\alpha q_\beta,
  \nonumber\\
  P_{4}^{\mu \nu}(P,q) &= \frac{4 Q^2 P \cdot q}{\Delta^2} \left( P_{5}^{\mu \nu} + P_{6}^{\mu \nu} - \frac{8 P \cdot q}{\Delta^3} \left( Q^2 P^{\mu} P^{\nu} + P^2 q^{\mu} q^{\nu} \right) \right),
  \nonumber\\
  P_{5}^{\mu \nu}(P,q) &= \frac{P \cdot q}{\Delta} \left( g^{\mu \nu} + \frac{4}{\Delta^2} \left( P^2 q^{\mu} q^{\nu} - Q^2 P^{\mu} P^{\nu} - P \cdot q \left( P^{\mu} q^{\nu} + P^{\nu} q^{\mu} \right) \right) \right),
  \nonumber\\
  P_{6}^{\mu \nu}(P,q) &= -\frac{2 P \cdot q}{\Delta^3} \left( 4 P \cdot q (P^{\mu} q^{\nu} + q^{\mu} P^{\nu}) + \frac{\Delta^2 - 8 P^2 Q^2}{Q^2} q^{\mu} q^{\nu} \right),
  \nonumber \\
  P_{7}^{\mu \nu}(P,q) &= -2 P_{6}^{\mu \nu} -\frac{8  P \cdot q}{\Delta^3} \left( P \cdot q (P^{\mu} q^{\nu} + q^{\mu} P^{\nu}) - 2 P^2 q^{\mu} q^{\nu} \right)\, .
\end{align}
The kinematic factor $\Delta$ is the Källén triangle function $\Delta \equiv \Delta((P+q)^2, P^2, -Q^2)$, where
\begin{equation}
    \Delta(a, b, c) = \sqrt{a^2 + b^2 + c^2 - 2(ab + ac + bc)}.
\end{equation}
By construction, the projectors satisfy the relation
\begin{equation}
    P_{i}^{\mu \nu}( P,q) \Delta W_{\mu \nu}= g_i, \quad \Delta W_{\mu \nu}=\frac{1}{2} (W_{\mu \nu }(-S)- W_{\mu \nu }(S))
\end{equation}
with
\begin{equation}
    S^{\mu}= \frac{2 P \cdot q}{\Delta} \left(P^{\mu}-\frac{P^2}{P \cdot q} q^\mu\right).
\end{equation}
The partonic projectors can be obtained from
\begin{equation}
    \hat{P}_{i, \, \rm QS}^{\mu \nu}=P_{i}^{\mu \nu}(k_1,q) , \quad  \hat{P}_{i, \, \rm GF}^{\mu \nu}=P_{i}^{\mu \nu}(p,q)
\end{equation}

It is important to emphasize that our calculation of the Wilson coefficients utilizes four-dimensional projectors for the structure functions. In conventional massless schemes (e.g., Ref.~\cite{Vogelsang_1996}), projectors are typically formulated in $D$ dimensions to regulate collinear divergences. However, in this framework, the parton masses provide a physical cutoff for such divergences, justifying the use of a four-dimensional approach for the projectors themselves. Conversely, we maintain a fully $D$-dimensional treatment of the phase space to regulate soft singularities. This ensures the consistent cancellation of infrared (IR) poles between real and virtual contributions, following the methodology established in Ref.~\cite{Kretzer:1998ju}.

\section{Virtual and soft coefficients}
\label{Appendix Virtual}

This section presents the calculations derived from the virtual correction illustrated in~\cref{fig:Vertex Correction}. To calculate the diagram, it is essential to recognize that the helicity eigenstates satisfy the Dirac equation \cite{Haber:1994pe}, i.e.,
\begin{equation}
    \not{k}_1 u (k_1, h)=  m_1 u (k_1, h), \quad \not{k}_1 v(k_1, h)=-m_1 v(k_1,h).
\end{equation}
Consequently, the vertex correction is the same as in the unpolarized case \cite{Kretzer:1998ju}. Specifically, the result is given by
\begin{equation}
\begin{aligned}
 \Gamma_0^{\mu} & =C_F \frac{\alpha_s}{4 \pi} \frac{1}{\Gamma(1-\epsilon)}\left(\frac{Q^2}{4 \pi \mu^2}\right)^{-\epsilon}\left(C_{0,-} \gamma^\mu L_5+C_{+} \gamma^\mu R_5\right. \\
& \left.+C_{1,-} m_2 k_1^\mu L_5+C_{1,+} m_1 k_1^\mu R_5+C_{q,-} m_2 q^\mu L_5+C_{q,+} m_1 q^\mu R_5\right)
\label{vertex correction analitical expression}
\end{aligned}
\end{equation}
with $ L_5=(V- A \gamma_5)$ and $R_5= (V+A \gamma_5)$. The coefficients in $D=4-2 \epsilon$ read
\begin{equation}
\begin{aligned}
&C_{0,-}=\frac{1}{\epsilon}\left(1+\Sigma_{++} I_1\right)+\left[\frac{\Delta^2}{2 Q^2}+\Sigma_{++}\left(1+\ln \left(\frac{Q^2}{\Delta}\right)\right)\right] I_1\\
& +\frac{1}{2} \ln \left(\frac{Q^2}{m_1^2}\right)+\frac{1}{2} \ln \left(\frac{Q^2}{m_2^2}\right)+\frac{m_2^2-m_1^2}{2 Q^2} \ln \left(\frac{m_1^2}{m_2^2}\right)+\frac{\Sigma_{++}}{\Delta} \\
& \times\left\{\frac{1}{2} \ln ^2\left|\frac{\Delta-\Sigma_{+-}}{2 Q^2}\right|+\frac{1}{2} \ln ^2\left|\frac{\Delta-\Sigma_{-+}}{2 Q^2}\right|-\frac{1}{2} \ln ^2\left|\frac{\Delta+\Sigma_{+-}}{2 Q^2}\right|-\frac{1}{2} \ln ^2\left|\frac{\Delta+\Sigma_{-+}}{2 Q^2}\right|\right. \\
& \left.-\operatorname{Li}_2\left(\frac{\Delta-\Sigma_{+-}}{2 \Delta}\right)-\operatorname{Li}_2\left(\frac{\Delta-\Sigma_{-+}}{2 \Delta}\right)+\mathrm{Li}_2\left(\frac{\Delta+\Sigma_{+-}}{2 \Delta}\right)+\mathrm{Li}_2\left(\frac{\Delta+\Sigma_{-+}}{2 \Delta}\right)\right\},\\
&C_{+}  =2 m_1 m_2 I_1, \\
&C_{1,-}  =-\frac{1}{Q^2}\left[\Sigma_{+-} I_1+\ln \left(\frac{m_1^2}{m_2^2}\right)\right], \\
&C_{1,+}  =-\frac{1}{Q^2}\left[\Sigma_{-+} I_1-\ln \left(\frac{m_1^2}{m_2^2}\right)\right], \\
&C_{q,-}  =\frac{1}{Q^4}\left[\left(\Delta^2-2 m_1^2 Q^2\right) I_1-2 Q^2+\Sigma_{+-} \ln \left(\frac{m_1^2}{m_2^2}\right)\right], \\
& C_{q,+}  =\frac{1}{Q^4}\left[\left(-\Delta^2+2 m_2^2 Q^2-\Sigma_{-+} Q^2\right) I_1+2 Q^2+\left(\Sigma_{-+}+Q^2\right) \ln \left(\frac{m_1^2}{m_2^2}\right)\right]
\end{aligned}
\label{VirtualCoefficients}
\end{equation}
with
\begin{equation}
I_1=\frac{1}{\Delta} \ln \left(\frac{\Sigma_{++}+\Delta}{\Sigma_{++}-\Delta}\right), \quad \Sigma_{\pm \pm} = Q^2 \pm m_2^2 \pm m_1^2. 
\end{equation}
The renormalized vertex is given by
\begin{equation}
    \Gamma_{\rm R}^{\mu}= \gamma^{\mu} L_5 (Z_1-1)+ \Gamma_{0}^{\mu}, 
\end{equation}
where
\begin{equation}
Z_1=1+C_F \frac{\alpha_s}{4 \pi} \frac{1}{\Gamma(1-\epsilon)}\left(\frac{Q^2}{4 \pi \mu^2}\right)^{-\epsilon}\left[-\frac{3}{\epsilon}-\frac{3}{2} \ln \left(\frac{Q^2}{m_1^2}\right)-\frac{3}{2} \ln \left(\frac{Q^2}{m_2^2}\right)-4\right].
\end{equation}
The only coefficient that changes is $C_{0,-}$, which is replaced by $C_{R,-}$ given by
\begin{equation}
\begin{aligned}
C_{R,-} & =-\frac{1}{\epsilon}\left(2-\Sigma_{++} I_1\right)+\left[\frac{\Delta^2}{2 Q^2}+\Sigma_{++}\left(1+\ln \left(\frac{Q^2}{\Delta}\right)\right)\right] I_1 \\
& +\frac{m_2^2-m_1^2}{2 Q^2} \ln \left(\frac{m_1^2}{m_2^2}\right)-\ln \left(\frac{Q^2}{m_1^2}\right)-\ln \left(\frac{Q^2}{m_2^2}\right)-4+\frac{\Sigma_{++}}{\Delta} \\
& \times\left\{\frac{1}{2} \ln ^2\left|\frac{\Delta-\Sigma_{+-}}{2 Q^2}\right|+\frac{1}{2} \ln ^2\left|\frac{\Delta-\Sigma_{-+}}{2 Q^2}\right|-\frac{1}{2} \ln ^2\left|\frac{\Delta+\Sigma_{+-}}{2 Q^2}\right|-\frac{1}{2} \ln ^2\left|\frac{\Delta+\Sigma_{-+}}{2 Q^2}\right|\right. \\
& \left.-\operatorname{Li}_2\left(\frac{\Delta-\Sigma_{+-}}{2 \Delta}\right)-\operatorname{Li}_2\left(\frac{\Delta-\Sigma_{-+}}{2 \Delta}\right)+\operatorname{Li}_2\left(\frac{\Delta+\Sigma_{+-}}{2 \Delta}\right)+\operatorname{Li}_2\left(\frac{\Delta+\Sigma_{-+}}{2 \Delta}\right)\right\}.
\end{aligned}
\label{CRminus}
\end{equation}

By combining the virtual correction with the LO contribution, we obtain the coefficients \( V_i \) and \( N_i \) reported in~\cref{Inclusive QS}, i.e.
\begin{align}
    V_1 =& C_{R-} + \frac{q_- \Sigma_{++} - q_+ 2 m_1 m_2}{q_+ \Sigma_{++} - q_- 2 m_1 m_2} C_{+}, \nonumber\\ 
    V_4 =& C_{R-} + \frac{1}{2} \left( m_2^2 C_{1,-} - m_1^2 \frac{\Sigma_{-+}}{\Sigma_{+-}} C_{1,+} \right)+ \frac{R_-}{R_+} \left( \frac{m_1 m_2}{2 } \left( C_{1, \,+}- \frac{\Sigma_{-+}}{\Sigma_{+-}} C_{1, -}   \right)+ C_+  \right), \nonumber\\ 
     V_5 =& C_{R-} + \frac{R_{-}}{R_{+}} C_+, \nonumber\\ 
     V_6 =& C_{R-} +   \frac{R_{-}}{R_{+}} C_+  +\frac{1}{2 \left( (m_2^2 - m_1^2) \Delta^2 + 4 m_1^2 Q^2 \Sigma_{-+}\right)} \times \bigg\{\nonumber \\ &+   \Sigma_{+-}^2 \left( m_2  C_{1,-}\left(m_2 \Sigma_{+-}- m_1 \Sigma_{-+} \frac{R_{-}}{R_+}\right) - m_1  C_{1,+} \left(m_1 \Sigma_{-+}- m_2 \Sigma_{+-} \frac{R_{-}}{R_{+}} \right) \right) \nonumber\\ &+ 4 m_1^2 Q^4 \left( m_2 C_{q,-}\left( 2m_2+ \frac{\Sigma_{++}}{m_1} \frac{R_-}{R_+}\right)+  C_{q,+} \left(\Sigma_{++}+ 2 m_1 m_2 \frac{R_-}{R_+}\right)\right) \bigg\}, \nonumber\\
V_7 =& C_{R-} + \frac{R_-}{R_+}\left(C_+ + C_{1,-} \frac{m_1 m_2}{2} \right) \nonumber\\
&+ \frac{1}{2} \Bigg\{ m_1 \left( m_1 + m_2 \frac{\Delta^2 - 2 m_1^2 Q^2}{\Delta^2 - Q^2 \Sigma_{++}} \frac{R_-}{R_+} \right) C_{1,+}  + \frac{1}{\Delta^2 - Q^2 \Sigma_{++}} \Big( m_2^2 (\Delta^2 - 2 m_1^2 Q^2) C_{1,-} \nonumber\\
&+ Q^2 \Big( m_1 \left( m_1 \Sigma_{-+} - m_2 \Sigma_{+-} \frac{R_-}{R_+} \right) C_{q,+}  - m_2 \left( m_2 \Sigma_{+-} - m_1 \Sigma_{-+} \frac{R_-}{R_+} \right) C_{q,-} \Big) \Big) \Bigg\} ,\nonumber\\
    N_1=&  \frac{1}{2\Delta}( q_+ \Sigma_{++}- q_- 2 m_1 m_2 ) ,\nonumber \\
    N_4=&  2R_+ \left(\frac{\Sigma_{+-}}{\Delta}\right)^3 ,\nonumber\\ 
    N_5 =& R_+ \left( \frac{\Sigma_{+-}}{\Delta}\right) ,\nonumber\\ 
    N_6 =& R_+\frac{ \Sigma_{+-}}{Q^2 \Delta^3} \Big( (m_2^2 - m_1^2) \Delta^2 + 4 m_1^2 Q^2 \Sigma_{-+} \Big), \nonumber\\ 
    N_7 =& - 2R_+ \frac{ \Sigma_{+-}^2}{Q^2 \Delta^3} \Big( \Delta^2 - Q^2 \Sigma_{++} \Big),
\end{align}
where $R_{-}= (A^\prime V- A V^\prime)/2$.
The pole is precisely canceled when taking into account the contribution arising from the phase-space integration of the real contribution, i.e.,
\begin{align}
    S_i &= \lim_{\hat{s}_{1} \rightarrow 0} \left( \frac{\Delta^2}{m_{2}^2 Q^2} \right)^{-\epsilon} \left( \frac{1}{\epsilon}\frac{\hat{s}_{1}^2}{m_{2}^2}\right) \int_{0}^{1} y^{-\epsilon} (1-y)^{-\epsilon} \left( \frac{m_{2}^2}{\hat{s}_1^2}+ \frac{m_{1}^2}{\hat{t}_{1}^2}+ \frac{\Sigma_{++}}{\hat{s}_1\hat{t}_1}       \right)  \, dy + \mathcal{O}(\epsilon)  \nonumber\\
    &= -\frac{1}{\epsilon}\left(-2+\Sigma_{++} I_1\right)+2+\frac{\Sigma_{++}}{\Delta}\left(\Delta I_1+\mathrm{Li}_2\left(\frac{2 \Delta}{\Delta-\Sigma_{++}}\right)-\mathrm{Li}_2\left(\frac{2 \Delta}{\Delta+\Sigma_{++}}\right)\right)+\mathcal{O}(\epsilon).
\end{align}

\acknowledgments

This work has been supported by the BMFTR under contract 05P24PMA. E.S. expresses sincere gratitude to Felix Hekhorn and Daniel de Florian for helpful suggestions and discussions.

\bibliography{polACOT_references}

\end{document}